\documentclass{aa}

\usepackage{graphicx}
\usepackage{txfonts}
\usepackage{xcolor}
\usepackage{adjustbox}
\usepackage{subcaption}
\usepackage{caption}
\usepackage{url}

\begin{document} 
 
   \title{Towards an astronomical use of new-generation geodetic observations}


   \subtitle{I. From the correlator to full-polarization images}

   \author{V. Pérez-Díez\inst{1,4},
    I. Martí-Vidal\inst{2,3},
    E. Albentosa-Ruiz\inst{2},
    J. González-García\inst{4},
    F. Jaron\inst{5,8},
    T. Savolainen\inst{6,7,8},
    M.~H.~Xu\inst{9,6},
    R. Bachiller\inst{1}}


   \institute{
   Observatorio Astron\'omico Nacional (OAN-IGN), Alfonso XII 3, 28014 Madrid, Spain
   \and Dpt. Astronomia i Astrof\'isica, Universitat de Val\`encia, C/ Dr. Moliner 50, 46120 Burjassot, Spain
   \and Observatori Astron\`omic, Universitat de Val\`encia, C/ Cat. Jos\'e Beltr\'an 2, 46980 Paterna, Spain
   \and Centro de Desarrollos Tecnológicos, Observatorio de Yebes (IGN), 19141 Yebes, Guadalajara, Spain
   \and Technische Universit\"at Wien, Wiedner Hauptstra\ss{}e 8-10, 1040 Wien, Austria
   \and Aalto University Mets\"ahovi Radio Observatory, Mets\"ahovintie 114, 02540 Kylm\"al\"a, Finland
   \and Aalto University Department of Electronics and Nanoengineering, PO Box 15500, 00076 Aalto, Finland
   \and Max-Planck-Institut f\"ur Radioastronomie, Auf dem H\"ugel 69, 53121 Bonn, Germany
   \and GFZ German Research Centre for Geosciences, 14473 Potsdam, Germany
   }

  \date{\today}

  \titlerunning{Wideband global calibration strategies for VGOS}
  \authorrunning{Pérez-Díez, Martí-Vidal, Albentosa-Ruiz et. al.}

 
  \abstract
   {The current algorithms used for the calibration and analysis of very long baseline interferometry (VLBI) networks that only use linear polarizers (as is the case of the VLBI Global Observing System, VGOS) do not properly account for instrumental and source-intrinsic polarimetry, which can cause errors in geodetic and astronomical products.}
   {We aim to develop a calibration pipeline for VLBI interferometers that observe in a basis of linear polarization, as is the case of VGOS. The products from this pipeline can be used to obtain valuable full-polarization astronomical information from the observed sources, and they can be used to potentially improve the geodetic results.}
   {We used the algorithm PolConvert to write the correlation products in a basis of circular polarization that is compatible with the standard VLBI calibration procedures. In addition to this, we implemented a wide-band global fringe-fitting algorithm that accounts for dispersive effects (ionospheric delay) and allows us to perform full-polarization imaging of all the observed sources, covering the whole frequency band of VGOS.}
   {We present the outcome of our pipeline applied to a global IVS VGOS epoch of observations and show example imaging results in total intensity and polarization. We also discuss issues encountered during the analysis and suggest points of improvement in the VGOS system for an optimum geodetic and astronomical exploitation of this interferometer.}
   {}


   \maketitle
%

\section{Introduction}

Traditionally, circular-polarization receivers have been commonly employed in very long baseline interferometry (VLBI) for various reasons. VLBI observations involve extremely long baselines, which introduce significant differences in the parallactic angles of the different antennas. Correcting for the effects of parallactic angle in the interferometric phase is easier while observing in a circular polarization basis (i.e., dividing the signal into right-hand, RCP, and left-hand, LCP, circular polarizations), since the effect of the parallactic-angle rotation is corrected for with just a deterministic phase applied to the visibilities. In contrast, if linear-polarization feeds are used in the VLBI observations, the gain differences (phases and amplitudes) between the polarization channels at each antenna (and for each frequency) must be known before the parallactic-angle correction can be applied. Therefore, the calibration and fringe-fitting strategy for linear-polarization feeds may become significantly more complex and computationally expensive because this precalibration is needed prior to the fringe fitting \citep[for a deeper discussion, see][]{PolConvert}.

However, circular-polarization receivers also have certain disadvantages. On the one hand, the additional hardware required for the conversion into circular polarization (e.g., quarter-wave plates at the receiver frontend) may add additional noise and degrade the polarization purity of the receiver, resulting in a higher instrumental polarization and a slightly lower signal-to-noise ratio (S/N). On the other hand, the fractional bandwidth available is narrower because the circular-polarization hardware-based converters are designed for an optimum performance at a specific frequency, which worsens their polarization purity as we depart from that frequency. Due to this last drawback, linear-polarization receivers may indeed be the most convenient choice for VLBI antennas with very wide band receivers (e.g., with fractional bandwidths of about 100\%). 

This is the case of the VLBI Global Observing System (VGOS), the next-generation geodetic VLBI system coordinated by the International VLBI Service for Geodesy and Astronomy (IVS)\footnote{\url{https://ivscc.gsfc.nasa.gov/index.html}}. In order to achieve the goal of 1 mm accuracy in station position, it requires extremely wide bandwidths to allow the  measurement of group delays with a precision of about a picosecond. In particular, VGOS demands the use of ultra-wideband receivers covering a broad frequency range from 2\,GHz to 14\,GHz, as was originally planned \citep{VGOSRef} and recently confirmed \citep{Niell2018}. 



Various algorithms have been proposed to calibrate the data from VGOS observations. The officially used algorithm for processing is found in \cite{HayCap} and is based on the use of the so-called pseudo-Stokes I (henceforth, pI), a quantity obtained by combining the four linear-polarization correlation products (i.e., $XX$, $XY$, $YX$, and $YY$), accounting for instrumental cross-polarization phase (not amplitude) bandpasses, and parallactic-angle differences between antennas. The main drawback of this method for its astronomical exploitation is that it only handles total intensity, so that it is not possible to obtain information about the polarized brightness distribution of the observed sources. Furthermore, deriving the epoch-wise instrumental cross-polarization phase bandpasses of all the antennas is usually an expensive procedure in the pI-based approach. It may take several hours of processing for a typical global VGOS observation. 

In order to address these drawbacks and find ways to improve the calibration and analysis of VGOS, a European project called EU-VGOS was initiated, as reported in \citet{EuVGOS}, and the current status has been reported in  \citet{EuVGOS23}. The EU-VGOS project decided to use the software PolConvert \citep{PolConvert} as a way to convert the VGOS data into a circular polarization basis just after correlation, in such a way that legacy VLBI calibration algorithms (all based on observations with circular-polarization receivers) could be used in VGOS. We refer to the conversion from linear into circular polarization using the tool PolConvert as \textit{polconversion}.

The algorithm PolConvert is able to postprocess VLBI data obtained with linear-polarization receivers (or a mixture of linear and circular polarization feeds at different antennas), writing the products as if circular polarization feeds had been used at all antennas. The immediate advantage of this software is the possibility to retrieve all four Stokes parameters (I, Q, U, and V) from the observed sources with a low instrumental polarization because the full cross-polarization bandpasses, including amplitude ratios, are accounted for in the conversion process. This algorithm has already been successfully applied to several types of observations \citep[e.g.,][]{EHTRef, LBARef2, M87_GMVA, LBARef, EVNRef}, ranging from millimeter and submillimeter-VLBI (e.g., the Event Horizon Telescope, EHT, and the Global Millimeter VLBI Array, GMVA) to centimeter-VLBI (e.g., the Long Baseline Array, LBA, and the European VLBI Network, EVN).

In this paper, we present a detailed description of the complete calibration process of the data obtained during a global VGOS 24 hr session by using an approach based on PolConvert. All steps are described, starting from correlation to polconversion with existing software. Then, we present our new global fringe-fitting algorithm and the innovative amplitude calibration and multifrequency imaging of sources at full polarization. The pipeline proposed in this paper has been developed with the objective of obtaining astronomical information from VGOS observations, in addition to the possibility of improving the geodetic products when the effects of the (polarized) structure of the observed sources are removed from the geodetic data \citep[e.g.,][]{XuStructure}. In forthcoming publications, we will analyze the polarization morphology (and frequency dependence) of the images presented here, as well as the geodetic products obtained from these data as compared to those obtained from the official VGOS calibration strategy (i.e., the algorithm based on pI).

In Sect. \ref{sec:Observations}, we summarize the observational setup (participating antennas, scheduling, frequency setup, correlation strategy, etc.). In Sect. \ref{sec:data_calibration}, we describe our pipeline, emphasizing the steps related to polconversion, global wide-band fringe fitting, and amplitude calibration. In Sect. \ref{sec:results}, we discuss the calibrated visibilities obtained from our pipeline in comparison to those from the official VGOS pipeline \citep{HayCap} and present (Sect. \ref{sec:resultsImages}) full-polarization images of a selection of sources. We also propose strategies (Sect. \ref{sec:resultsStrategies}) to help improve the performance of VGOS in future observations. Finally, in Sect. \ref{sec:conclusions}, we summarize our conclusions.

\section{Observations}
\label{sec:Observations}

The observations presented here correspond to the IVS experiment with code VO2187, which was observed on 6-7 July 2022. The eight participating antennas (with their code names) were Goddard (GS), Ishioka (IS), Kokee (K2), McDonald (MG), the twin Onsala telescopes (OE and OW), Westford (WF), and Yebes (YJ). The antenna coordinates used in the correlation are listed in Table \ref{tab:antennas}.

The total recorded bandwidth was 1\,GHz, divided into 32 spectral windows (spw) of 32\,MHz each. The spw are distributed across a wide frequency coverage, ranging between 3\,GHz and 11\,GHz, and are arranged in subsets of four bands (i.e., eight spw each). 
The bands, centered around 3.25, 5.5, 6.75, and 10.5\,GHz, respectively, are labeled A, B, C, and D, following the convention of the VGOS community. 

\begin{table*}[]
    \centering
    \caption{Antennas used in experiment VO2187, together with their reference geocenter coordinates as used in the DiFX correlation.}
    \begin{tabular}{l|c|r|r|r}
    \multicolumn{2}{c|}{ Antenna } & \multicolumn{3}{c}{GC Coords (m).} \\ 
    \hline
    IVS Component Name & Code & X & Y & Z \\
    \hline 
    Goddard Geophysical and Astronomical Observatory      & GS & $1130729.877$   &  $-4831245.972$  &  $3994228.300$ \\
    Ishioka VLBI Station         & IS & $-3959636.203$  &  $3296825.448$   &  $3747042.571$  \\
    Kokee Park Geophysical Observatory & K2 & $-5543831.745$  &  $-2054585.590$  &  $2387828.974$  \\
    McDonald   Geodetic Observatory     & MG & $-1330788.462$  &  $-5328106.593$  &  $3236427.492$ \\
    Onsala Space Observatory East     & OE & $3370889.298$   &  $711571.199$    &  $5349692.048$ \\
    Onsala Space Observatory West     & OW & $3370946.779$   &  $711534.507$    &  $5349660.925$  \\
    Westford Antenna, Haystack Observatory & WF & $1492206.223$   &  $-4458130.552$  &  $4296015.629$ \\
    IGN Yebes Observatory  & YJ & $4848831.021$   &  $-261629.388$   &  $4122976.576$  \\
    
    \hline
    \end{tabular}
    \label{tab:antennas}
\end{table*}

Phase-calibration tones equally spaced in frequency (hereafter, phase-cal tones) were inserted at each station, covering the full band in intervals of 5\,MHz, which ensured a total of six to seven phase-cal tone detections per spw for a robust determination of the instrumental delay and phase at each frequency. The exception was the Yebes antenna (YJ), for which the phase-cal tones were spaced in intervals of 10\,MHz (i.e., only three to four usable tone detections per spw). This limitation at YJ may degrade the quality of its instrumental phase calibration, as discussed in Sect. \ref{sec:instrumentalPhases}.

The correlation was processed at the MPIfR in Bonn using DiFX-2.5.4 \citep{DiFXref} in full-polarization mode, that is, the four polarization correlations, $HH$, $HV$, $VH$, and $VV$, were produced. A total of 160 spectral channels, with a correlator accumulation time of 1\,s, were generated for each spw. DiFX refers the correlation times to the geocenter, which allows the computation of closure quantities from the correlated outputs \citep[e.g.,][]{TMSref}. These closures are not affected by antenna gain effects (i.e., they only depend on the brightness distributions of the sources), which allows the imaging of VGOS data if a global fringe-fitting algorithm \citep{CottonSchwab} is used (see Sect. \ref{sec:GFF}).

A total of 74 sources were observed (radio-loud AGN; in particular, blazars), spanning an overall observing time of 24 hours in total. The sources were observed in short interleaved scans with a typical duration of 30\,s. The number of scans was different for each source, ranging from just one scan (source 0847-120) up to 66 scans (source 1803+784). In Table \ref{tab:sources}, we list the sources on which more observing time was spent, and for which full-polarization images were generated (see Sect. \ref{sec:resultsImages}).

\begin{table}
    \centering
    \caption{Sources most commonly observed in experiment VO2187.}
    \begin{tabular}{r | r | c}
    Source & $N_{obs}$ & $N_{scan}$ \\
    \hline
    1849+670 & 1004 & 60 \\ 
    2229+695 &  987 & 61 \\ 
    1803+784 &  836 & 66 \\ 
    0059+581 &  692 & 59 \\ 
    0613+570 &  625 & 46 \\ 
    3C418  & 532 & 56 \\ 
    0955+476 &  392 & 52 \\ 
    3C274 &  203 & 39 \\ 
    DA426 &  177 & 43 \\ 
    OJ287 &  168 & 37 \\ 
    \hline
    \end{tabular}
    \label{tab:sources}
    \tablefoot{Sources are ordered by the number of observations ($N_{obs}$, which is the sum of the number of baselines per scan, over all scans). $N_{scan}$ is the number of scans. Sources 3C274, DA426 and OJ287 (which should be further down in the full source list) are included in this table for scientific interest.\\}
\end{table}



\section{Data calibration}
\label{sec:data_calibration}


\subsection{PolConvert cross-polarization bandpass}
\label{sec:polconvert}


\begin{figure*}
    \includegraphics[width=0.98\textwidth]{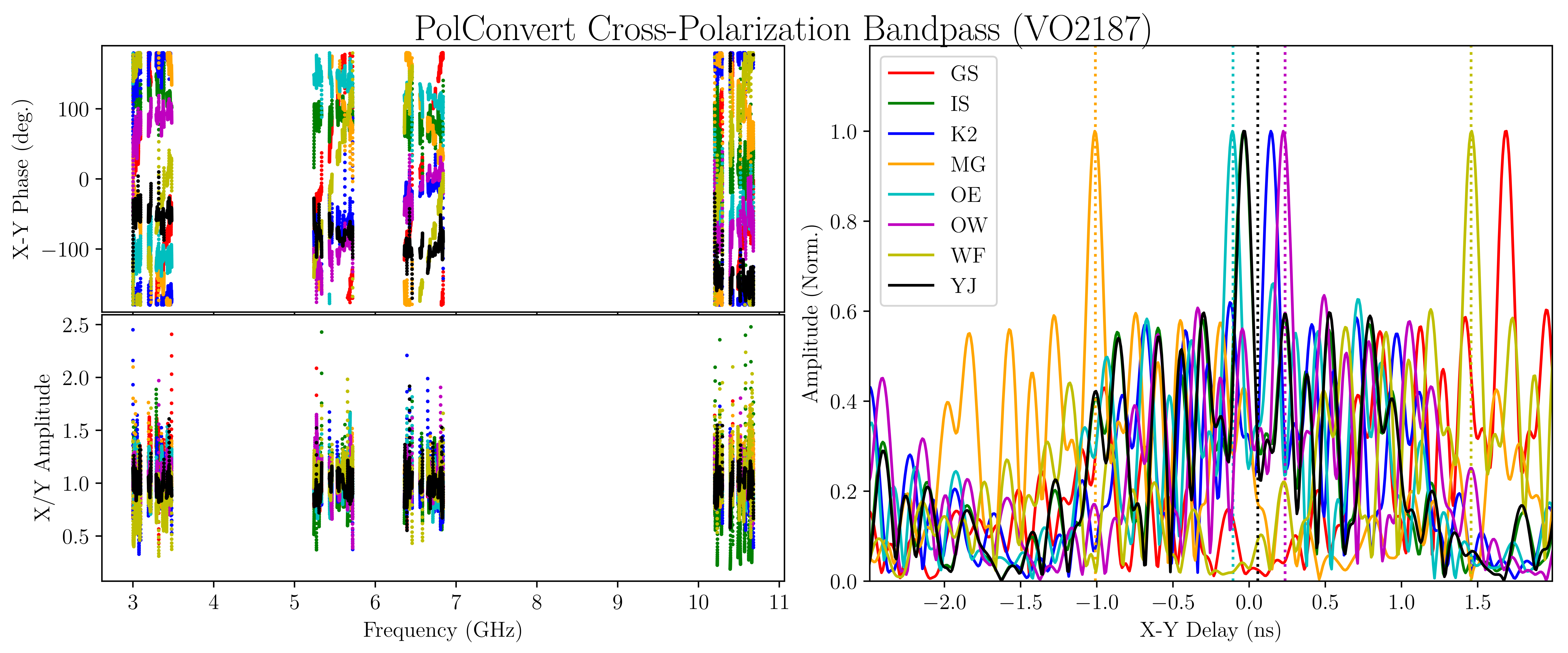}
    \caption{Cross-polarization bandpasses for experiment VO2187 (left; phases at top; amplitudes at the bottom), as estimated with PolConvert (see text). Multiband cross-polarization delays (right; computed from the values shown at the left). The dotted lines (same colors) mark the cross-polarization multiband delays obtained 63 days later from experiment ER2201 \citep{Jaron23}.}
    \label{fig:XPBandpass}
\end{figure*}

PolConvert \citep{polconvertzenodo} was originally designed to work on ALMA-VLBI observations. Depending on the observing frequency, the ALMA receivers have different feed orientations concerning the antenna mounts. Historically, PolConvert therefore used the generic nomenclature $X$ and $Y$ for the polarization channels (which should not be confused with the labels of the Cartesian geocentric coordinates) instead of the alternative $H$ and $V$, which are only rigorously correct for the special case in which $Y$ is oriented parallel to the exact azimuth axis of the antenna. For consistency with the nomenclature used in the PolConvert code and all its driver scripts, we therefore henceforth identify $H$ and $V$ with $X$ and $Y$, respectively. 

The conversion from linear to circular polarization is primarily affected by the cross-polarization ($X/Y$) bandpass of each antenna in phase and amplitude. These quantities can be estimated in PolConvert using the algorithm called global cross-polarization fringe-fitting (GCPFF) \citep{PolConvert}.

To apply this algorithm, the first step is to select one or more calibrator scans. In order to choose them, we performed a coarse baseline-based preliminary fringe fitting on a selected set of spw for all sources in the four correlation products ($XX$, $YY$, $XY$, and $YX$), and we calculated the S/R of the fringes. In this way, we selected suitable calibration scans with a good S/R. It is preferable to have either a long calibration scan and/or a set of scans covering some parallactic-angle changes in the antennas, which are needed to break a 180-degree global phase ambiguity in the $X/Y$ bandpass. This ambiguity is equivalent to exchanging $R$ and $L$ polarizations at all antennas. In short, we required a source that was bright, was observed with a sufficient parallactic-angle coverage in any antenna, and had (preferably) a low fractional linear polarization.

In the process of estimating the $X/Y$ bandpasses, PolConvert applies a priori phase information computed from the difference of the phases of the extracted phase-cal tones between $X$ and $Y$ for each spw and at each integration time. This difference was fit as a cross-polarization tone delay for each spw, which was then added to the $X/Y$ bandpass to be solved. 


These cross-polarization bandpasses were then applied to the remaining scans of the experiment because any drift in the $X/Y$ phase differences was tracked by the phase-cal tones that are used by PolConvert in all scans. In essence, the quantities estimated by the GCPFF can be interpreted as related to the difference in the optical path of the $X$ and $Y$ polarization channels from the receiver frontend until the point at which the phase-cal tones are injected (located after the polarization splitter in the waveguides). In our procedure, we therefore assumed that the part of the hardware that causes the GCPFF cross-polarization gains was stable at the timescale of the duration of the experiment.

In Fig. \ref{fig:XPBandpass}, we show the GCPFF cross-polarization bandpasses obtained with PolConvert using the first two scans of VO2187 (observations of sources 0059$+$581 and 1741$-$038). The phases and amplitudes of the $X/Y$ bandpasses are shown in the left panel (each antenna with a different color). The bandpasses in delay space are shown in the right panel, where the dotted lines mark the delays reported for another VGOS epoch \citep{Jaron23} for the subset of antennas that also participated in VO2187.
Several conclusions can be drawn from this figure. First, the $X/Y$ phases for each antenna can be well connected across the whole band (from 3 to 11\,GHz), without jumps or discontinuities. This indicates that the $X/Y$ bandpass effects introduced by the receiver components before the injection of the phase-cal tones can be basically described by a single delay between the polarization channels. All $X/Y$ amplitude ratios were around unity, which indicates a similar electronic gain for the two polarization channels at all antennas. There are some exceptions, however. For instance, IS in band D shows a low $X/Y$ ratio, as does WF in band A.

The quality of the alignment of the $X/Y$ phases across the VGOS band appears clear in the fringe plots shown in Fig. \ref{fig:XPBandpass} (right), which show clean peaks that also remain stable (differences smaller than 10\,ps) across a time baseline of more than two months when we compare the peak positions to those reported in \cite{Jaron23} for epoch ER2201 (observed 63 days after VO2187). The exception is the YJ antenna, for which a shift of about 100\,ps is seen between the $X/Y$ delay in VO2187 and ER2201. The shift can be explained by changes in the phase-cal system at YJ between the two epochs because some amplifiers at the end of the analog signal chain were replaced due to failure.

\subsection{PolConverted fringes}

After the GCPFF cross-polarization bandpasses were obtained as described in the previous section, we executed PolConvert to apply them (together with the polarization difference of the phase-cal tones) to the whole experiment.

One way to assess the quality of the polconversion is to compare the fringe amplitudes in the parallel-hand correlations (i.e., $RR$ and $LL$) to those of the cross-hand correlations (i.e., $RL$ and $LR$). In particular, the former should be remarkably higher than the latter under normal conditions (i.e., as long as the fractional polarization of the sources is low and their polarized structures are not much more compact than the Stokes I brightness distribution). This condition should hold for all sources at all observing times.

\begin{figure*}
    \includegraphics[width=0.515\textwidth]{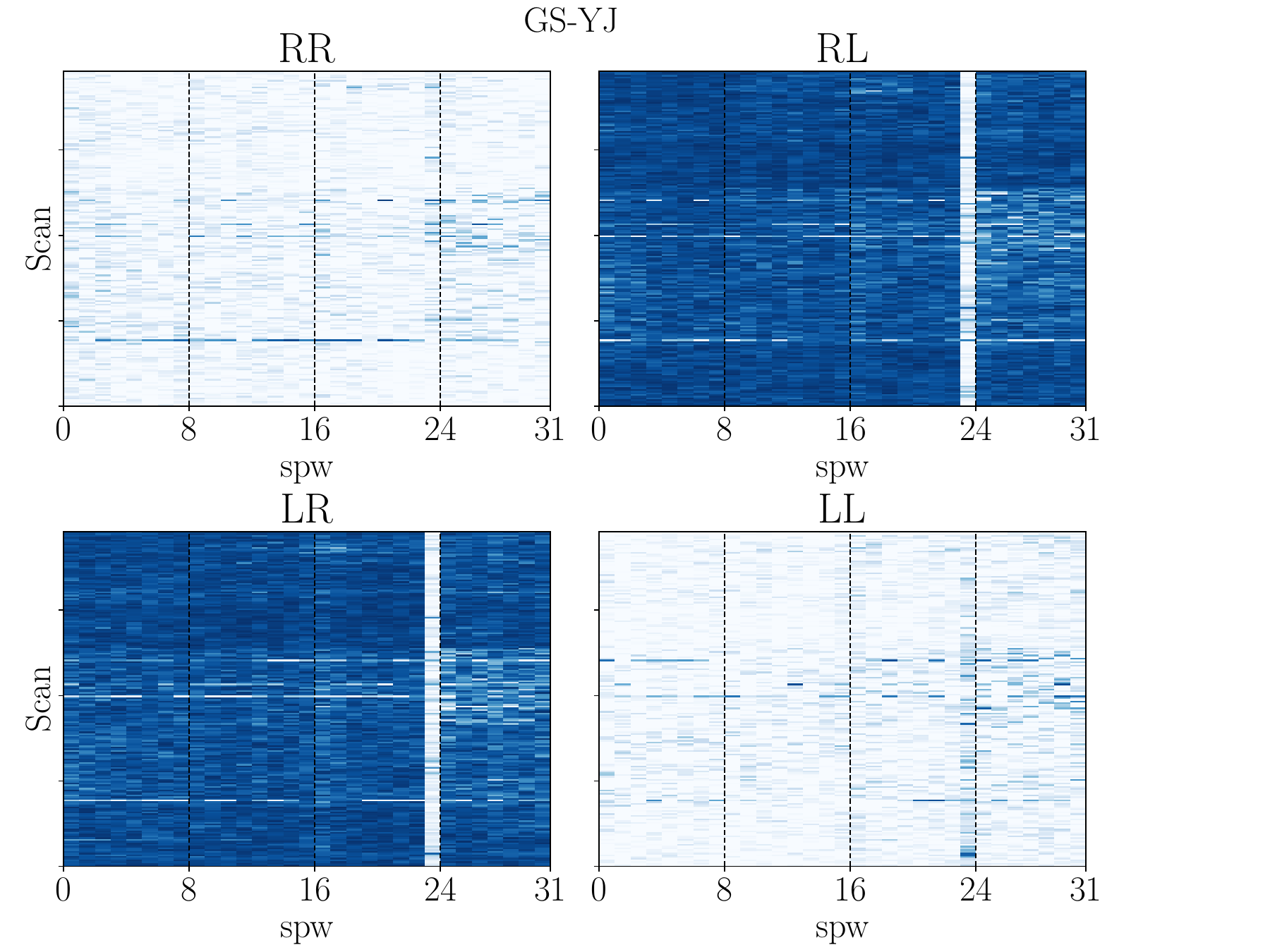}
    \hspace{-7mm}
    \includegraphics[width=0.515\textwidth]{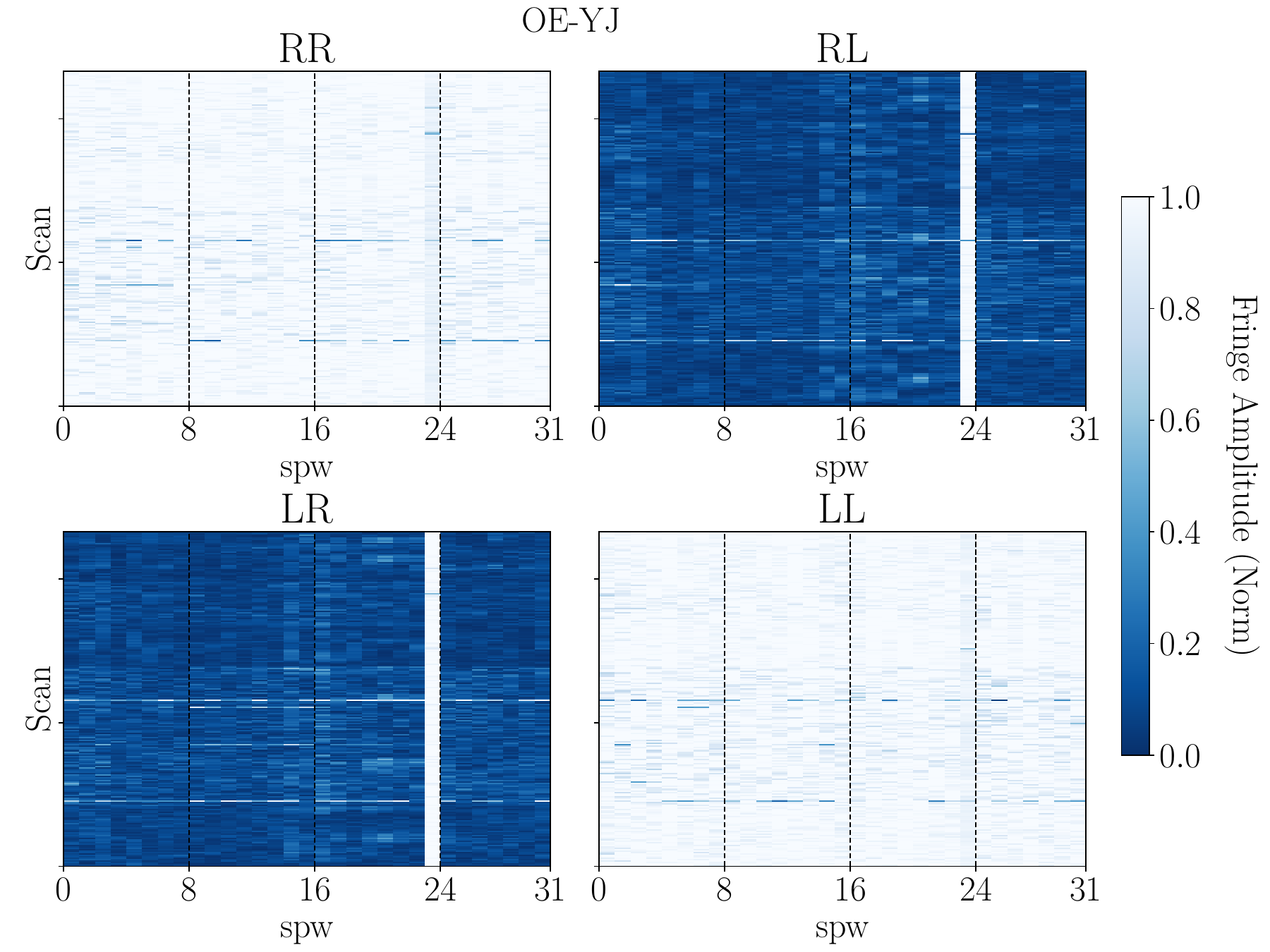}
    \caption{Fringe peaks for a selection of baselines as a function of spw (horizontal axis) and scan (vertical axis), and for the set of sources listed in Table \ref{tab:sources}. All four correlation products are shown. Left, baseline GS to YJ. Right, baseline OE to YJ.}
    \label{fig:FringePeak}
\end{figure*}

In Fig. \ref{fig:FringePeak}, we show the fringe amplitude peaks of baselines GS-YJ and OE-YJ in all spw (and all four polconverted correlation products) for all the scans of the seven most frequently observed sources (see Table \ref{tab:sources}). The different scans (a total of 197 and 234 for GS-YJ and OE-YJ, respectively) are arranged on the vertical axes of the plots, and the color palette for each scan is normalized to its peak. For all spw, the parallel-hand amplitudes of all these scans are significantly higher than the cross-hand amplitudes. The only exception is spw 23, which corresponds to the highest frequency of band C. In this spw, similar amplitudes are obtained in all four correlation products, which indicates that the polconversion has not worked properly at that particular frequency window. This failure may be related to a degraded quality in the phase-cal tones at YJ (see Sect. \ref{sec:instrumentalPhases}).  

From Fig. \ref{fig:FringePeak}, a clear conclusion can be drawn: The polconversion has produced much higher fringe amplitudes in the parallel-hand products of different sources at times separated by several hours, covering a large region of the sky with very distinct parallactic angles. Therefore, the polconversion is not only restricted to the calibration scans, but can be extrapolated to the whole experiment. In the next section, we discuss how the polconverted phases behave under feed rotations with respect to the sky frame, which is a solid complementary proof of the successful conversion of the whole experiment into a circular polarization basis (see Appendix \ref{sec:EVPA} for more information about absolute EVPA calibration).

\subsection{Instrumental phase calibration}
\label{sec:instrumentalPhases}

As explained in Sec. \ref{sec:polconvert}, PolConvert takes the difference in phase-cal tone phases between the $X$ and $Y$ polarization channels (i.e., $X/Y$ phase-cals) into account when it estimates the cross-polarization gains (using the GCPFF algorithm) and also when it converts the data. This instrumental difference between polarizations was added to the cross-polarization bandpasses estimated with the GCPFF algorithm at each integration time. Therefore, because the $X/Y$ tone difference is already accounted for in the polconversion, the instrumental phases affecting the polconverted $RR$ and $LL$ visibilities are described by exactly the same phase-cal phases, which correspond to those in the $Y$ polarization channel (i.e., the reference polarization used in the GCPFF).

These remaining instrumental phases were estimated in a way similar to the so-called multitone mode of the fourfit program, which is the option used in the official VGOS calibration pipeline\footnote{\url{https://www.haystack.mit.edu/wp-content/uploads/2020/07/docs_hops_011_multitone_phasecal.pdf}}. The multitone mode estimates a tone delay, $\tau_{pc}$, from the phase-cals, $\phi_i$, found in each spw, by obtaining the peak of the tone fringe. Then, the instrumental tone phase, $\phi_{pc}$, is computed as the average of the tone residual phases centered at the reference spw frequency, $\nu_0$, when the tone delay has been subtracted. The equation used is

\begin{equation}
    \phi_{pc} = \left< \phi_i - 2\pi\tau_{pc} (\nu_i - \nu_0)\right>.
    \label{eq:pcal}
\end{equation}

\begin{figure*}[ht!]
  \centering
  \includegraphics[width=0.9\textwidth]{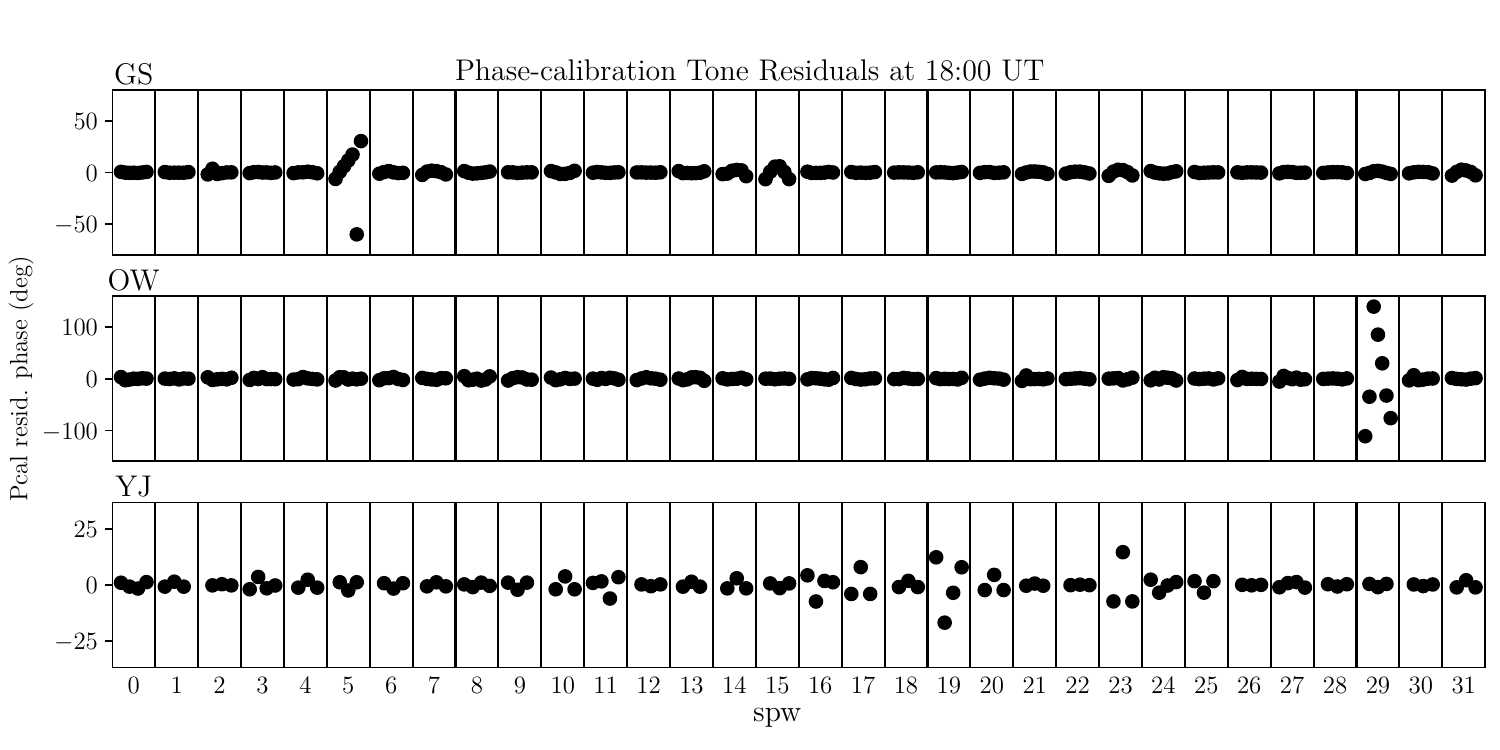}
  \caption{Residual phase-cal tones after subtraction of the estimated instrumental delays, without removal of the bad entries. Clear outliers are seen (e.g., for GS, OW, and YJ in spw 5, 29, and 19, respectively). However, it is not possible to identify some of the outliers at YJ (e.g., in spw 17 and 23; see text).}
  \label{fig:phasecals}
\end{figure*}

In Fig. \ref{fig:phasecals}, we show the residual phase-cal tone phases of the scan starting at 18\,UT for three representative antennas and after subtraction of the tone phases estimated using Eq. \ref{eq:pcal}. No outlier tones were removed. Some of the outliers are clearly seen, for instance, at spw 5 in GS or spw 29 at OW. Outlier tones result in incorrect estimates of the tone delays, $\tau_{pc}$, which in turn produce biased estimates of the instrumental tone phases, $\phi_{pc}$. Because several (6$-$7) tones are available within each spw, removing the outliers in these cases allowed us to retrieve the correct tone phases. However, for cases such as spw 23 of YJ, where only three tones are available for the whole spw, it is not possible to identify which tone is the actual outlier. This prevents a correct calibration of the tone delay, and as a consequence, the instrumental tone phase. This limitation clearly degrades the instrumental phase calibration at YJ for all the spw showing a behavior similar to that of spw 23.


   

\subsection{Manual phase alignment across the VGOS band}
\label{sec:ManualPhasecal}

After the phases were corrected using the phase-cal tones, as described in the previous section, we expected the visibility phases to be properly aligned across all the spw. However, this was not the case and was partially due, for instance, to bandpass effects from the receiver components prior to the injection of the phase-cal tones. In practice, the additional correction needed to properly align the phases across all the spw (quantities known as additive phases) was computed from the observation of a calibrator source. In the VLBI jargon, this procedure is known as manual phase-cal.

The ionosphere over each station may introduce strong effects in the visibility phases across a band as wide as that of VGOS. If the ionosphere is not accounted for in the manual phase-cal, strong biases in the estimate of dispersive and nondispersive delays from the global fringe fitting (see next section) may affect the final geodesy products.

\begin{figure*}[]
  \centering
  \begin{adjustbox}{trim={150pt} {10pt} {150pt} {0pt}, clip, width=1.01\textwidth}
    \includegraphics{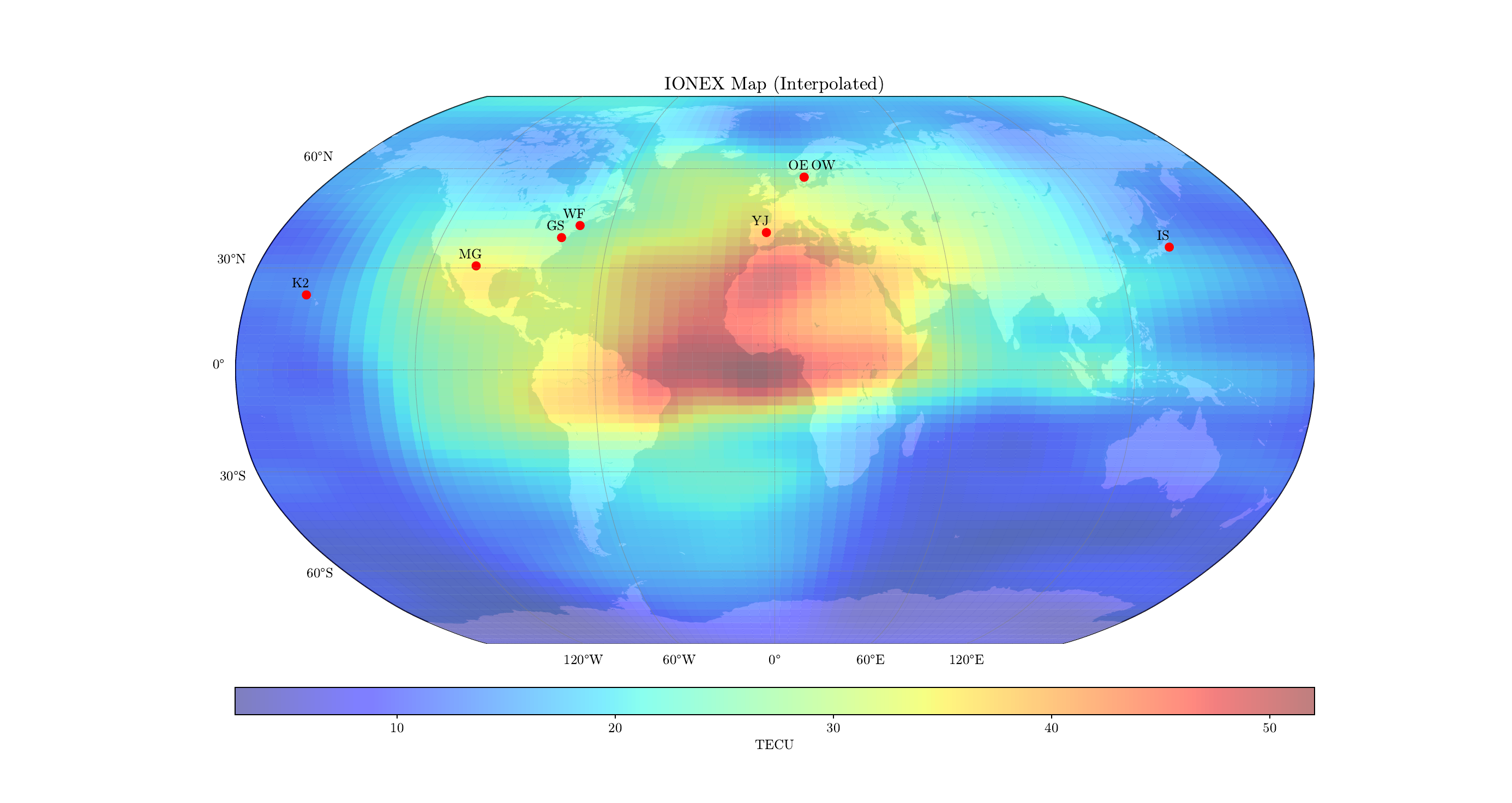}
  \end{adjustbox}
  \caption{Interpolated TEC map for scan 1749 (observed at 15.30\,UT), obtained using the global ionosphere map (IONEX) from the NASA Jet Propulsion Laboratory. The interpolation algorithm is similar to the one implemented in the AIPS task TECOR.}
  \label{fig:IONEX}
\end{figure*}

In order to minimize the contamination from the ionosphere in our manual phase-cal (and in the visibilities, in general), we used estimates of the total electron content (TEC) along the line of sight of each telescope, computed from GNSS IONEX models. In particular, we employed the algorithm PyPhases (included in PolConvert), which is based on the TECOR task implemented in the Astronomical Image Processing System (AIPS)\footnote{http://www.aips.nrao.edu}. Our algorithm reads ionospheric models produced by Jet Propulsion Laboratory (JPL) \footnote{https://cddis.nasa.gov/archive/gnss/products/ionex} that provide maps with the TEC at each point on the Earth's surface every 2 hours, assuming a thin ionospheric layer. A simple geometric model of the ionosphere was then employed to calculate the vertical TEC at the point at which the signal passes the thin layer. Then, spherical trigonometry was used to map this vertical TEC to the line of sight. Finally, we eliminated the ionospheric contribution from the observed phases at each integration time according to the following equation:

\begin{equation}
\label{eq:tec}
\phi_{obs} = \phi_0 + 2\pi \left(\tau_{gr}(\nu - \nu_0) + \dot{\tau}(t - t_0) \right) + \frac{\kappa \cdot \text{dTEC}}{\nu},
\end{equation}

\noindent where $\phi_{obs}$ is the observed phase, $\phi_0$ is the fringe phase, $\tau_{gr}$ is the nondispersive delay, $\nu$ is the frequency, $\nu_0$ is the fringe reference frequency, $\dot{\tau}$ is the fringe rate, $t$ is the visibility time, $t_0$ is the fringe reference time, and dTEC is the difference of the line-of-sight TEC (sTEC) between the two antennas of the baseline ($\kappa = 40.3 \, m^3s^{-2}$ is a constant, relating the TEC and the nondispersive delay).
In Fig. \ref{fig:IONEX}, we show the IONEX TEC model for the scan observed at 15:30\,UT, together with the geographical distribution of the antennas. In this scan, the ionosphere looks clearly more active for the antennas in Europe and in the continental US, where values of a few dozen TEC units (TECU) may affect their observations.

After the bulk of the ionosphere effects was removed from the data, we performed an ordinary global fringe fitting to each individual spw of the scan selected for the manual phase-cal with the IONEX model. From this GFF, we obtained the phase corrections that according to the manual phase-cal approach, definitely align the phases among all the spw in all the scans of the experiment, allowing us to execute the final wide-band global fringe Fitting, as we explain in detail in the following section.

It is worth noting that any small difference (in the manual-phase-cal calibration scan) between the true ionosphere and the prediction from the IONEX model is inevitably absorbed in the estimated phase corrections. In any case, this effect is propagated to the remaining experiment as a constant (and small) residual ionosphere offset, and it is expected to affect the geodetic observables and not the dTEC.

In the official VGOS approach, a manual phase-cal is also performed, where many scans that have high S/R fringes on all four polarization products and the dTEC solution for each polarization product is estimated to be lower than 1 TEC unit \citep{VGOSDataProcessingManual}. However, in addition to not using any a priori information to subtract the dispersive delay, the main difference is that they use a nonglobal approach for the fringe fitting, while in our procedure, we performed a global fringe fitting, as we explain in the next section.

As a final remark, in our manual phase-cal, we impose a multiband delay (MBD) at each antenna that equals the median of its single-band delays (SBD) for all the spw. In this way, we can combine MBD and SBD in our WBGFF into one single delay quantity (i.e., a delay that describes the frequency dependence of phases within the spw, as well as across the whole VGOS band). This strategy is discussed in more detail in the next section.

\subsection{Wide-band global fringe fitting}
\label{sec:GFF}



A global fringe-fitting algorithm solves for the instrumental gains using all the data available for each scan, but parameterizing the gain information in an antenna-based manner. By fitting the data globally, we optimally use all available information in a scan while minimizing the parameter space (which is proportional to the number of antennas, whereas the data size increases quadratically with that number). In addition, the global gain solutions are more robust against source-instrinsic effects than the baseline-based solutions. By globalizing the solutions, the closure phases \citep[quantities that encode information related to the brightness distribution of the observed sources, e.g. ][]{TMSref} remain unchanged in the data (i.e., not in the antenna gains) during the calibration, which allows us to perform image deconvolution. The effect of minimizing all the baseline phases, while keeping the phase closures unchanged, effectively concentrates the source brightness toward the image center. Baseline-based solutions, on the other hand, absorb all the effects related to the source structures (including closure phases) into their fitted quantities.  

To perform a global fringe fitting and be able to maintain the closure phases, all the visibilities at each scan must refer to the same time point, in this case, the moment in which the wavefront reaches the geocenter. However, in geodesy, each visibility refers to the moment in which the wavefront reaches the reference antenna at each baseline. In this way, the position of the antennas can be known without performing other calculations, which is why a nonglobal approach is used in geodetic fringe fitting.

In VLBI observations, the antenna gains in an ordinary fringe fitting are modeled using three quantities: instrumental phases (discussed in the previous subsections), nondispersive group delays that are mainly due to the array geometry and atmosphere, and fringe rates. The group delays can be measured either within the frequency channels of a single spw (these are the single-band delays, SBDs) or from the phases among different spws (these are the multiband delays, MBDs). In the official approach followed for VGOS, SBDs and MBDs are fit independently. The former are taken from the stacking of the delay fringes across all the spw. In our case, however (and becaise SBD and MBD are indeed two ways of measuring the same physical quantity: the delay between signals arriving at the two antennas), we performed the manual phase-cal (see previous section) in such a way that SBDs and MBDs were well aligned (i.e., the median phase slope within the spw was aligned to the phase slope across the whole VGOS band). In this way, we removed one of the fitting parameters in our WBGFF and only solved for the delay (i.e., MBD and SBD were combined). This method has some advantages compared to the approach followed by the official IVS calibration pipeline. On the one hand, it uses all the spectral channels of each spw in the fitting, whereas HOPS\footnote{\url{https://www.haystack.mit.edu/haystack-observatory-postprocessing-system-hops}} collapses all the frequency information of each spw into one single value when it estimates the MBD \citep{VGOSDataProcessingManual}. In our case, by using all the spectral information in the fit of the MBD, we effectively multiplied the MBD fringes by the Fourier transform of a 32 \,MHz boxcar function (i.e., the width of each spw, by means of the convolution theorem), hence decreasing the amplitudes of the MBD sidelobes and ambiguities. On the other hand, the WBGFF has only one delay parameter to model the SBDs and the MBDs. This decreases the number of degrees of freedom in the fitting.

In addition to this, if the fractional bandwidth is wide enough to detect dispersive effects due to the ionosphere, another parameter has to be introduced in the fringe fitting for each antenna: a dispersive delay, which depends on frequency as $\propto \nu^{-2}$. 

When fitting for dispersive and nondispersive delays simultaneously using Eq. \ref{eq:tec}, both parameters are usually highly coupled (i.e., the post-fit covariance matrix has high nondiagonal values), which may introduce biases in the estimate of the multiband nondispersive delay (i.e., the most important geodetic observable in VGOS). 

The WBGFF algorithm we developed simultaneously fits dispersive and nondispersive terms and tries to minimize the problems related to the parameter coupling. Unlike the official approach \citep{HayCap}, where fringe fitting is performed separately for each baseline, we chose to perform the fit globally, using antenna-based quantities as parameters. 

In this fitting process, we determine not only the phase and group delay, but also the residual ionosphere. After subtracting the IONEX priors (using the same approach as described in Sect. \ref{sec:ManualPhasecal}), the bulk of the ionosphere effects are removed from the data, so that we just adjust for the residual dTEC (i.e., the difference between the true ionosphere and that estimated from the IONEX model). 

Our WBGFF is implemented in the PyPhases script, which is part of the current distribution of PolConvert\footnote{https://github.com/marti-vidal-i/PolConvert}. In addition to globalizing the solutions, it also determines possible TEC ambiguities (i.e., high dTEC values that deform the fringes so that a sidelobe becomes higher than the main peak; see Fig. \ref{fig:TEC-AMBG}). We summarize the procedure followed by PyPhases in the following lines. For each scan, the following steps are taken:

\begin{figure*}
    \centering
    \includegraphics[width=18cm]{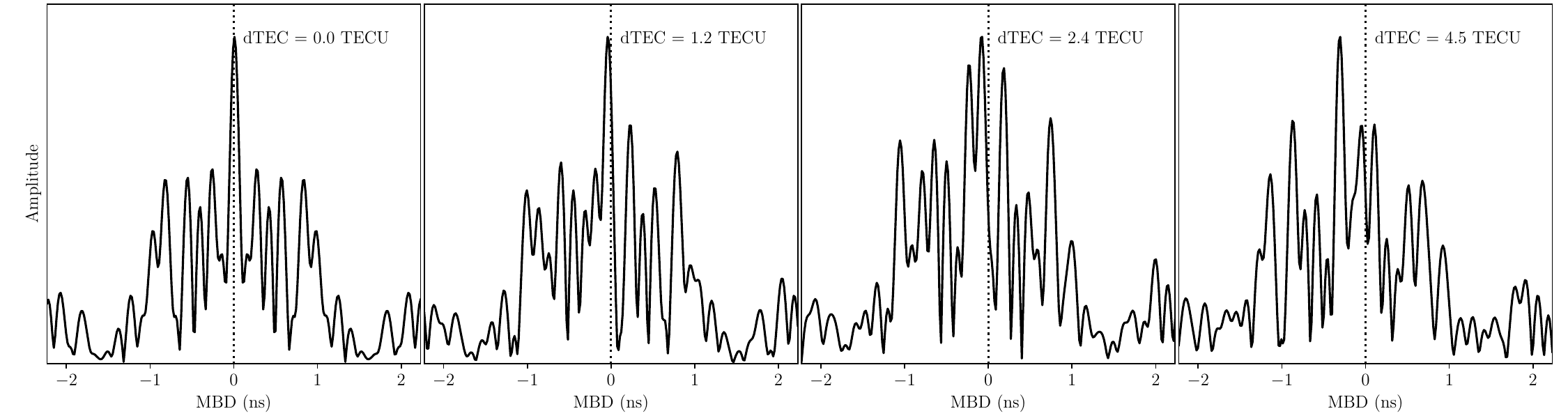}
    \caption{Simulated multiband delay fringes for the VGOS frequency coverage, assuming a point source without noise. Different dTEC values were added to each panel. The fringe peak with null dTEC is marked with a dotted line. The ambiguity introduced for dTEC = 4.5\,TECU is clear (i.e., the first sidelobe at the left becomes higher than the main fringe lobe).}
    \label{fig:TEC-AMBG}
\end{figure*}

\begin{enumerate}

\item Perform an ordinary GFF using a fringe model without dispersive terms (i.e., solving only for $\phi_0$, $\tau_{gr}$ and $\dot{\tau}$ in Eq. \ref{eq:tec}). From this fit, we obtain the so-called effective delay, $\tau_{ef} = \tau_{gr}$, which corresponds to the fringe peaks in the data. We note that, at this point, the fringe peaks are contaminated by the residual ionosphere.
\vspace{2mm}

\item Carry out a coarse exploration of TEC ambiguities by adding/subtracting TECU and comparing the S/R of the resulting GFF solutions. The solution with the highest S/R determines the TEC ambiguity.
\vspace{2mm}

\item Solve for the dispersive component using a modified version of Eq. \ref{eq:tec},

\begin{equation}
\label{eq:taum}
\phi = \phi'_0 + 2\pi\left(\tau_{ef} \cdot \nu + \dot{\tau}(t - t_0)\right) + \frac{\kappa \cdot \text{dTEC}}{\nu} \left(1 + \frac{\nu^2}{\nu_{ef}^2}\right),
\end{equation}

\noindent where $\nu_{ef} = \frac{1}{\sum_i \nu_i^{-1}}$. In this new equation, $\dot{\tau}$ is fixed to the value found in step 1, and $\tau_{ef}$ is highly constrained to a small window around the value found in that step.
\vspace{2mm}

\item Recover $\tau_{gr}$ by just replacing

\begin{equation}
    \tau_{gr} = \tau_{ef} + \frac{\kappa \cdot \mathrm{dTEC}}{\nu_{ef}^2}.
\end{equation}

\end{enumerate}



Through simulations, we verified that this method accurately estimates the residual dTEC as long as it remains below 5\,TECU (a condition that can be relaxed by using a wider dTEC window in step 2). 
With these steps, the phases would be completely calibrated, having decoupled the dispersive delay caused by the ionosphere, eliminating the instrumental phases and obtaining the necessary nondispersive group delay for geodesy.

The philosophy behind Eq. \ref{eq:taum} is that $\tau_{ef}$ has been designed to be equal (or very close) to the fringe peak when there is a small contamination from dispersive (ionosphere) delays. Hence, an ordinary (nondispersive) GFF is able to provide a good estimate of $\tau_{ef}$, which can then be tightly constrained in the process of solving for the dispersive components in step 3. In this step, the model would basically solve for the nonlinear component of the phase spectrum (i.e., the curvature in the phase model, which is only due to the ionosphere), while keeping the effective delay constrained.

On the other hand, in the HOPS procedure, they created a dTEC grid within a window of possible values and performed the fringe fitting within this grid, estimating the dTEC value as the value whose fringe had a higher S/R. By repeating this for each baseline, this makes the convergence slower than in our procedure, where dTEC is another parameter of the minimization.

\begin{figure*}
  \centering
  \includegraphics[width=0.98\textwidth]{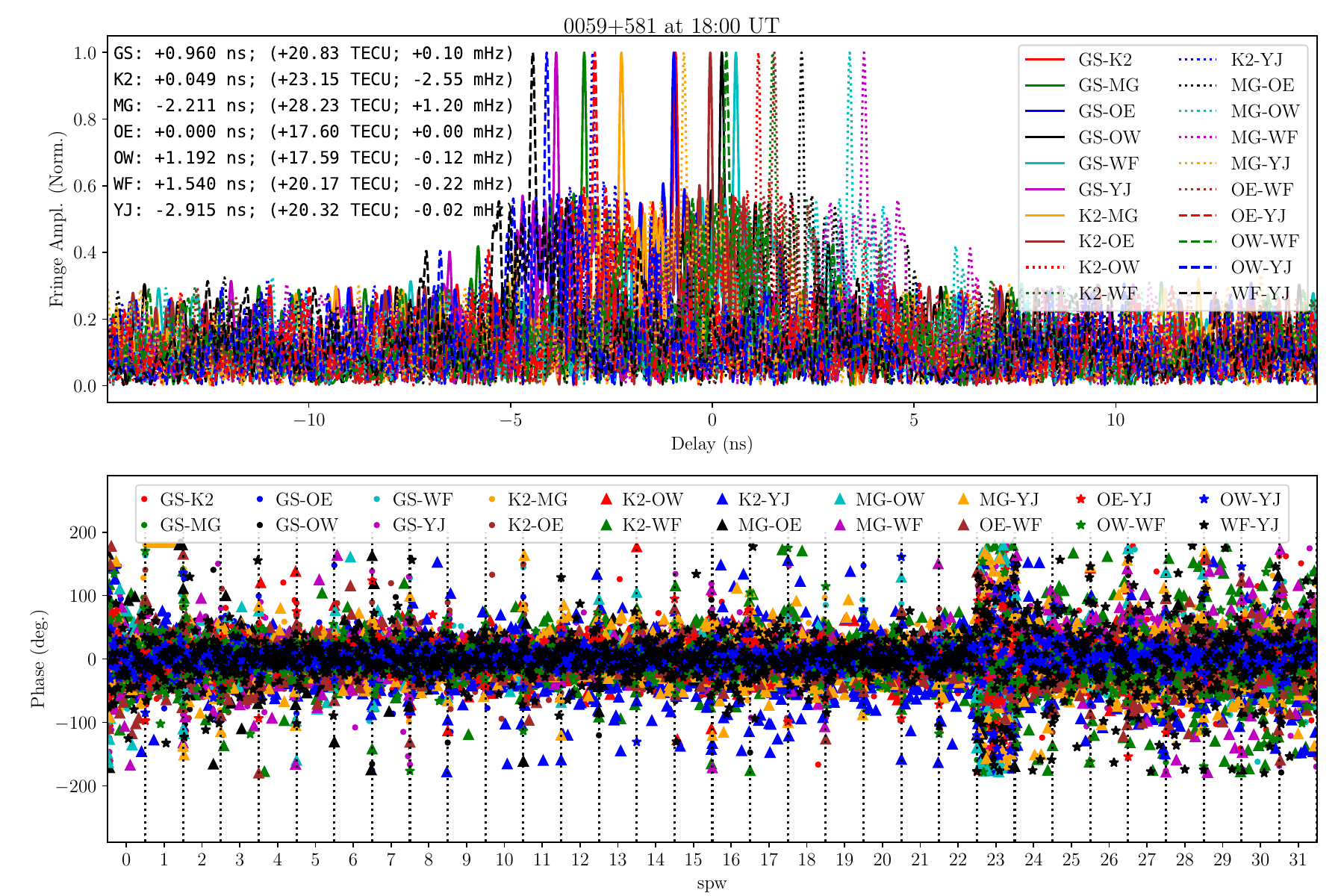}
  \caption{
Result of the WBGFF for the first scan of our experiment (taken at 18\,UT). Top: Fringe amplitudes in delay space after ionospheric dispersion is removed. The values of the antenna-based gains from the GFF are shown in the left panel (TECU is the full ionospheric contribution, including the prior from the IONEX map). The delay values are shown taking OE as the reference antenna. Bottom: Visibility phases after the complete fringe calibration, shown as a function of frequency (ordered by spw). The IS antenna did not participate in this scan.
}
  \label{fig:GFF}
\end{figure*}

This modular WBGFF implementation (i.e., starting with an ordinary GFF, followed by a fit of the phase-curvature terms) is very efficient (a factor of several times faster than the HOPS implementation for VGOS, although the use of different programming languages could also affect the benchmarks). In Fig. \ref{fig:GFF}, we show the WBGFF results on the first scan of the experiment, which we find representative of the typical fringe solutions. All fringes in the delay space shown in the top panel (different colors for different baselines) were fit using the antenna-based quantities listed in the left part of the panel (multiband delay, TEC, and fringe rate). The phase spectra for all the spw after applying the WBGFF calibration are shown in the bottom panel. No evidence of neither significant residual phases or multiband or single-band delays are seen, which indicates a successful calibration (see Appendix \ref{CASAComparison} for a comparison with CASA Fringe Fitting).

\begin{figure*}
\sidecaption
  \includegraphics[width=12cm]{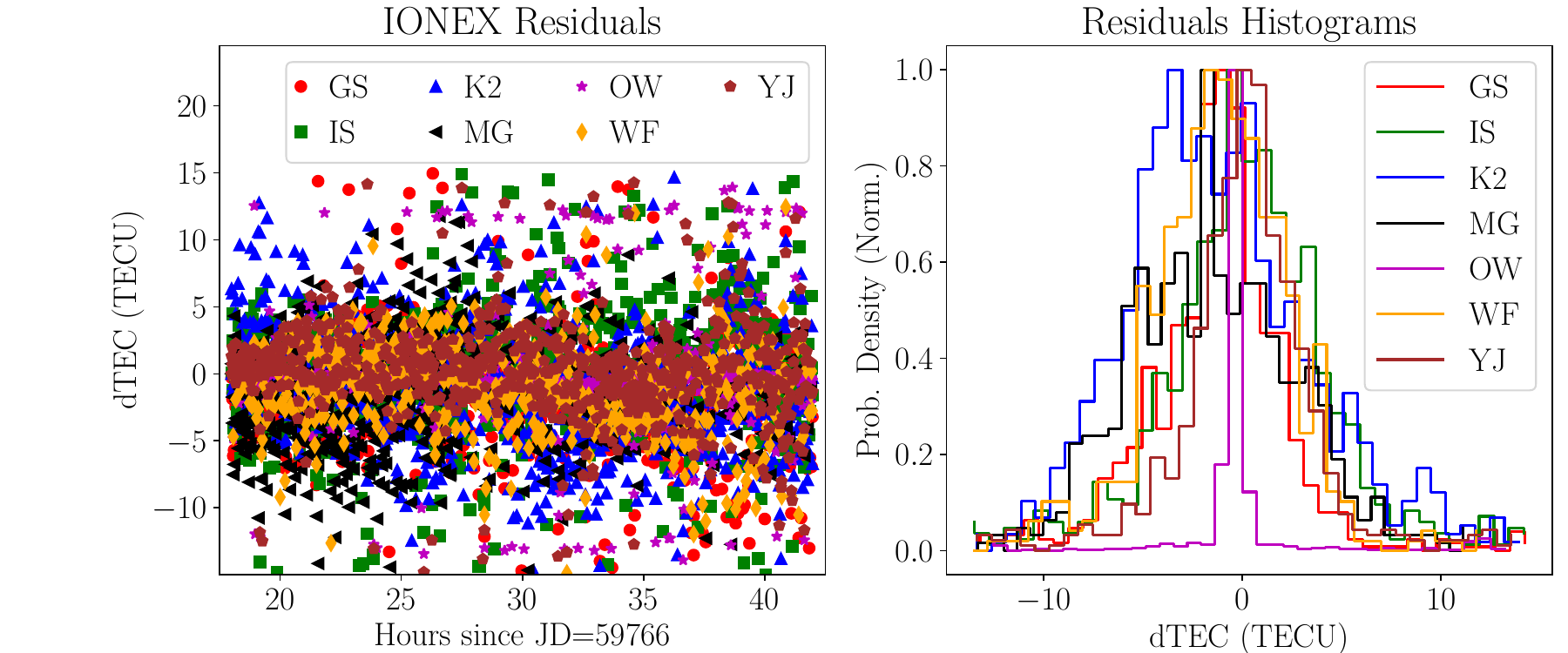}
  \caption{Residual dTEC (left; i.e., GFF estimates minus IONEX priors), referenced to antenna OE. Histogram distributions of the dTEC residuals (right), referenced to OE and normalized to the peak. }
  \label{fig:TECResid}
\end{figure*}

Regarding the estimates of the dTEC from our WBGFF, the residual quantities found for the whole experiment, referenced to antenna OE, are shown in Fig. \ref{fig:TECResid}. In the left panel, the residuals are shown as a function of time, with different symbols for each antenna. The right panel shows the histogram distributions of the antenna-based dTEC residuals for the whole experiment. Most of the residual dTEC falls within a window of -10$-$10\,TECU, depending on the baseline length. The dTEC of OW is a remarkable case, which takes typical values below 1\,TECU. This is an expected result because the very short baseline between the twin Onsala telescopes forces a very similar ionosphere for the two telescopes. In any case, this result also gives a good argument in favor of a global approach for the fringe fitting: the very well constrained ionosphere over OW (when OE is taken as reference) is used to constrain the fit of all the other antennas, which also share baselines to the two Onsala twin stations.

Another point worth noting in Fig. \ref{fig:TECResid} (right panel) is that the distribution of the residual dTEC of K2 peaks at a value clearly different than zero (around $-2$ to $-4$ TECU). This offset might be indicative of possible systematics in the IONEX model for a region above the Pacific Ocean.



\subsection{Amplitude calibration}

With our WBGFF, we fully calibrated the phases. The amplitude remained to be calibrated, for which we needed to calibrate the VGOS stations and retrieved the flux densities on each baseline. To compute the flux density $S_{i,j}$ of a baseline $i-j$, 
\begin{equation}
	S_{i,j} = A \sqrt{SEFD_i \cdot SEFD_j},
\end{equation}
where $A$ is the correlation amplitude, we need the station system equivalent flux densities (SEFDs), which were computed using the system temperatures ($\mathrm{T_{sys}}$), the DPFU (degrees per flux unit) and the antenna gains $g(z)$ (where $z$ is the elevation),
\begin{equation}
	SEFD = \frac{T_{sys}}{DPFU\cdot g(z)}.
\end{equation}
However, not all VGOS stations provide the full information required for the calibration. The IS and WF stations only provide SEFD and $\mathrm{T_{sys}}$ in every band at the start and end of each session, but do not measure $\mathrm{T_{sys}}$ nor the gain curve during observations.

After computing the SEFD and the flux density observation by observation for the sites with full information, we estimated the SEFD of the sites with medium information at each observation elevation, using a generic tipping curve. Then, we adjusted the SEFDs to line up the fluxes with those of sites with full information. After we estimated the flux density as a function of the uv distance, we performed an amplitude self-calibration using this function as a model.

\begin{figure*}
    \includegraphics[width=0.49\textwidth]{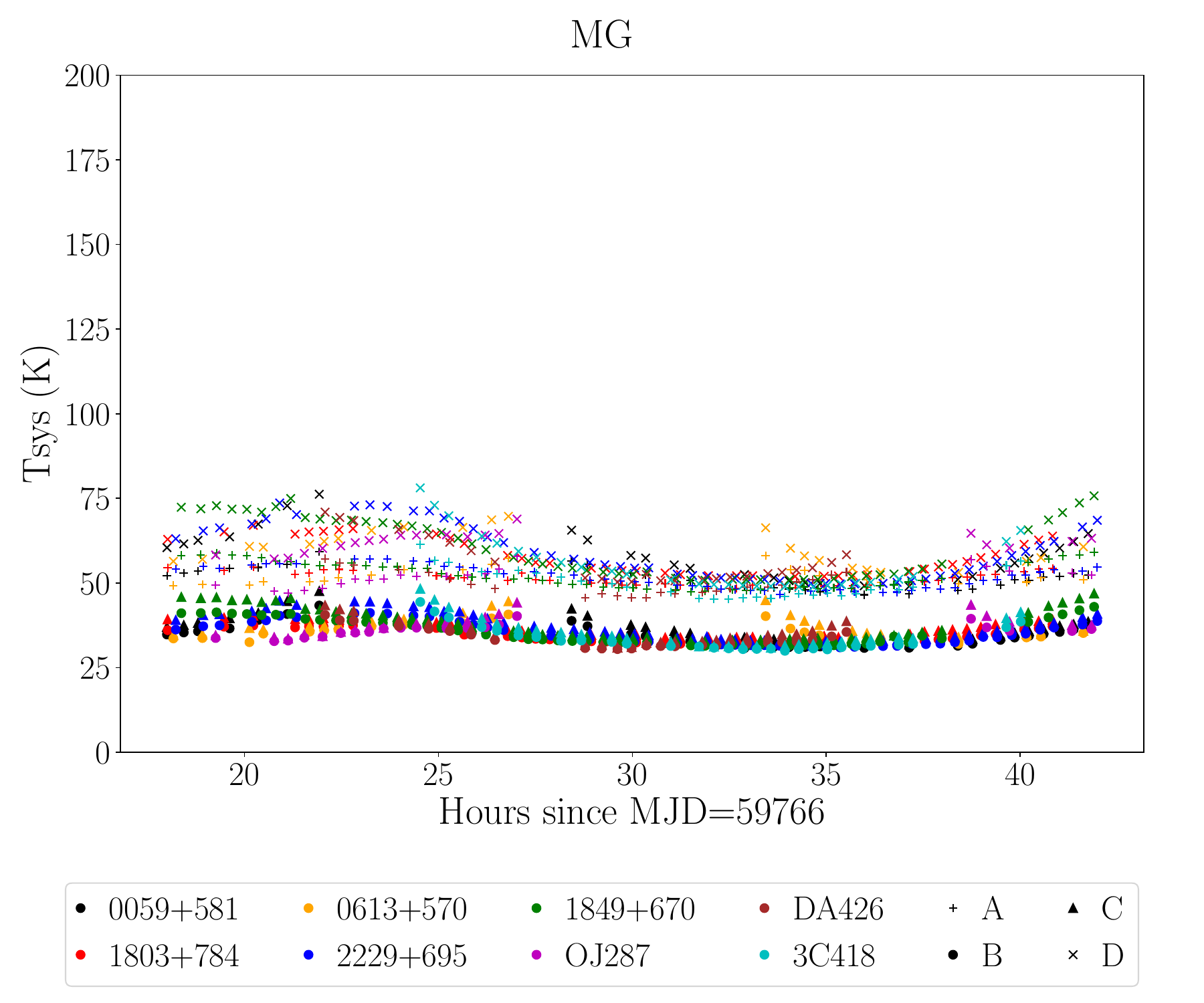}
    \includegraphics[width=0.49\textwidth]{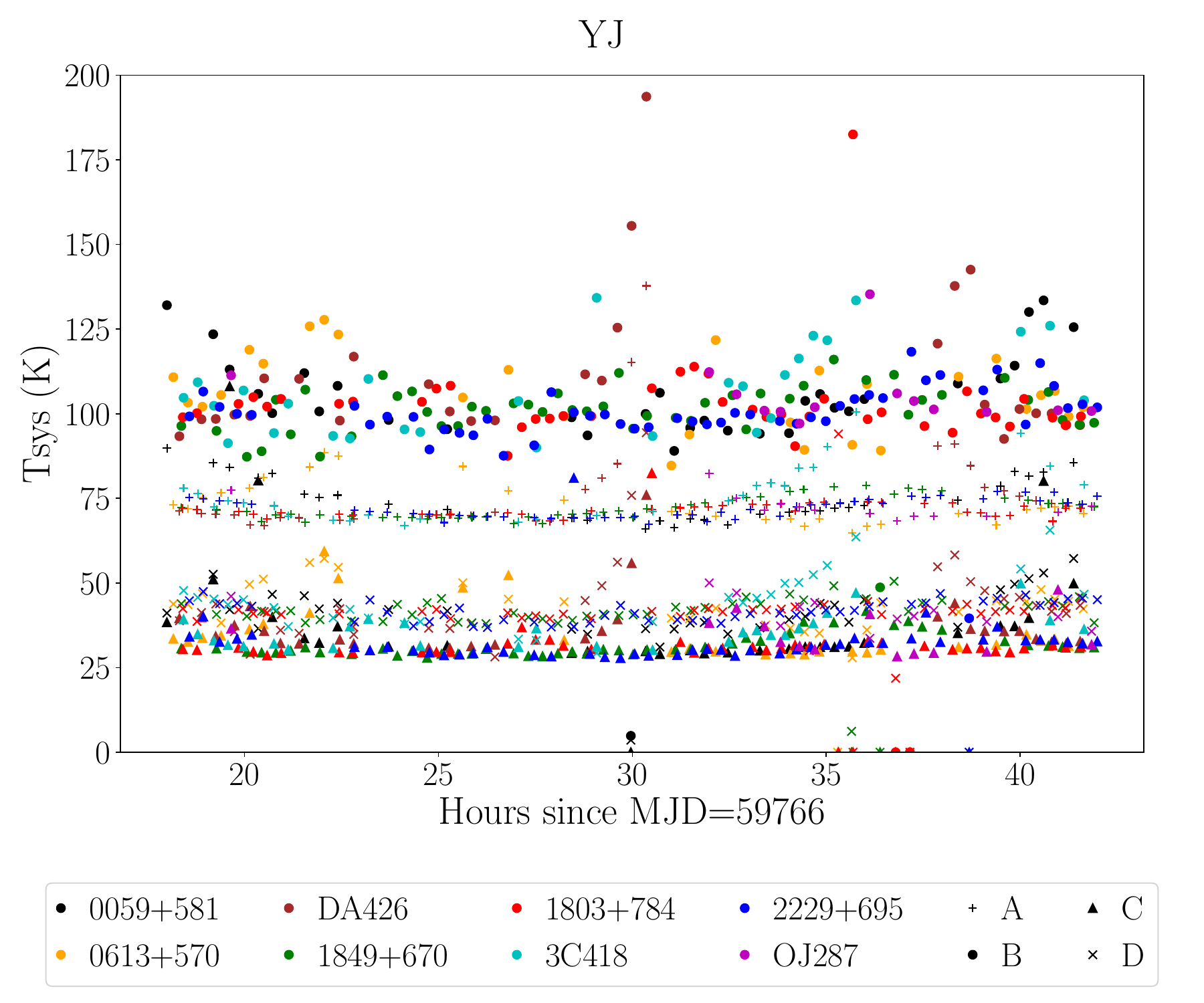}
    \caption{System temperatures ($T_{sys}$) for a selection of sources (different colors), averaged for each band (different symbols). For antenna MG (left). For antenna YJ (right).}
    \label{fig:TsysPlot}
\end{figure*}

The system temperatures of each VGOS antenna (with the exception of IS and WF, which do not provide this information) were used indirectly through the flux densities, which were estimated with the information provided by each station. In Fig. \ref{fig:TsysPlot}, we show the $\mathrm{T_{sys}}$ of two antennas, YJ and MG. MG shows hints of gain elevation effects, proving the need to characterize the gain to compute the SEFDs and estimate the baseline flux densities. The shape of the YJ station graph indicates the flatness of the gain curve of the ring-focus shaped Europe-based antennas.

\section{Discussion}
\label{sec:results}

\subsection{Comparison with the Haystack pipeline}


\begin{figure*}
\sidecaption
  \includegraphics[width=12.9782cm]{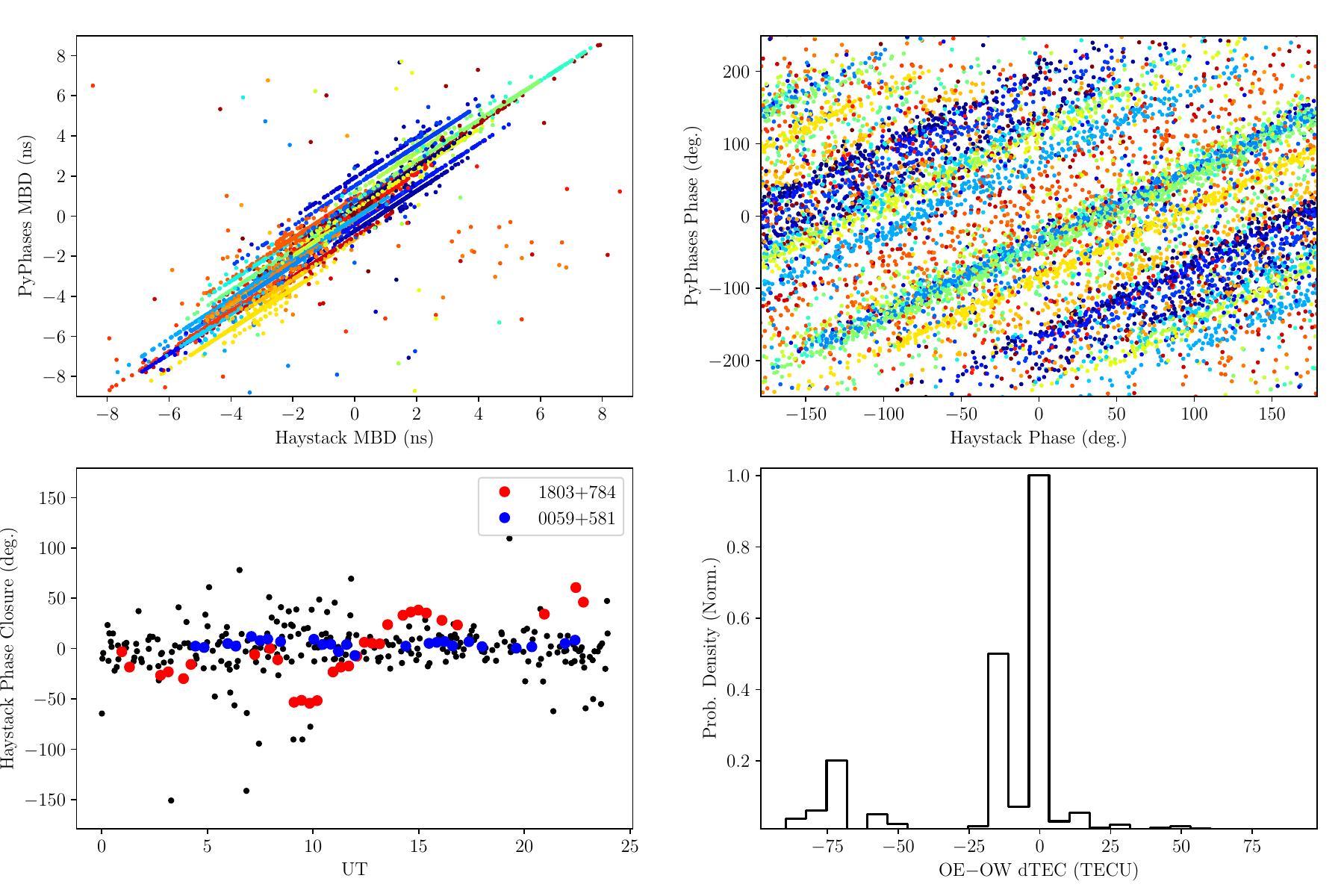}
  \caption{Comparison of fringe-fitted quantities between \texttt{fourfit} (following the Haystack pipeline) and PyPhases. Top: Correlation plots for the multiband delays (left) and phases (right). Baselines are shown with different colors. Bottom: Closure phases from \texttt{fourfit}, avoiding the OE-OW baseline (left), and histogram of dTEC differences between OE and OW (also from \texttt{fourfit}).}
  \label{fig:HaystackVsGFF}
\end{figure*}

 The antenna-based GFF solutions from PyPhases can easily be rewritten as baseline-based quantities by forming the differences between the quantities of the antennas of each baseline. The delays (both dispersive and nondispersive) can then be compared to the baseline-based solutions officially released for this VGOS epoch by the IVS, which were generated with the HOPS \texttt{fourfit} program following the control file generated by the Haystack VGOS pipeline \footnote{https://www.haystack.mit.edu}
 \texttt{vgoscf\_generate.py}, in its version on 8 September 2022. 

We expect a correlation between PyPhases and Haystack results in the form of a tight linear trend with baseline-dependent offsets, directly related to the different methods followed for the manual phase calibration. This is seen for the multiband delays and fringe phases (Fig. \ref{fig:HaystackVsGFF}, top panels). However, the baseline-based phase solutions from \texttt{fourfit} contain nonzero closure phases (same figure, bottom left panel), which are exactly zero (by construction) for the antenna-based PyPhases solutions.

Regarding the differential dispersive delay between the twin Onsala telescopes, OE and OW, in Fig. \ref{fig:HaystackVsGFF} (bottom right panel) we show the histogram from the \texttt{fourfit} Haystack pipeline. Even though the two stations should have very similar TEC corrections (because they are basically cospatial at scales of the ionosphere structure function) we see clear deviations from zero, which may be related to a degraded fringe quality of the OE-OW baseline, likely due to common RFI, cross-talk between the two stations, and correlation of their phase-cal tones (which can be partially mitigated by using notch filters). In our pipeline, we did not taken the OE-OW baseline data into account to avoid these drawbacks. However, because we used a global fringe fitting, we were able to obtain the dTEC independently and obtained better results with a dTEC value close to 0. For comparison, the equivalent histogram from PyPhases is shown in Fig. \ref{fig:TECResid} (magenta line in the right panel).

A more detailed comparison between the baseline-based Haystack (i.e., \texttt{fourfit}) results and PyPhases (with an emphasis on the effects in the geodetic analysis) is beyond the scope of this paper and will be published elsewhere. The good agreement seen between delays and phases of \texttt{fourfit} and PyPhases are a good confirmation of the reliability of our global fringe fitter.

\subsection{Calibrated visibilities}

\begin{figure*}
    \begin{minipage}{0.35\textwidth}
        \centering
        \includegraphics[width=\linewidth]{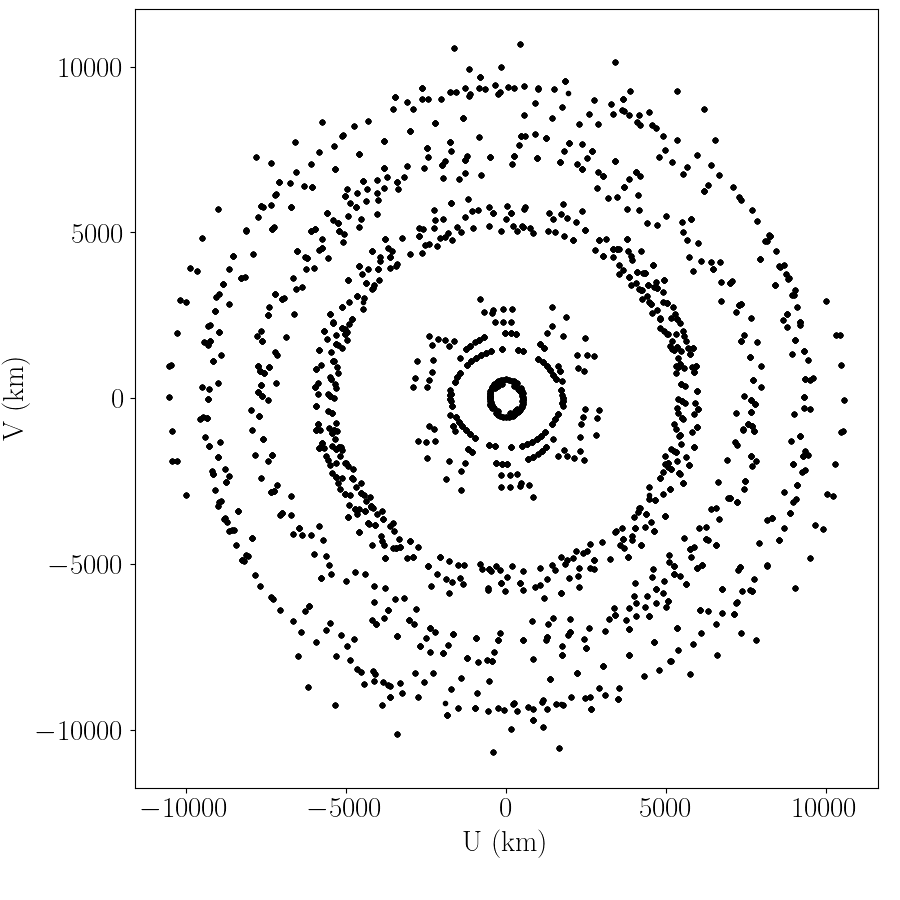}
    \end{minipage}
    \begin{minipage}{0.65\textwidth}
        \centering
        \includegraphics[width=\linewidth]{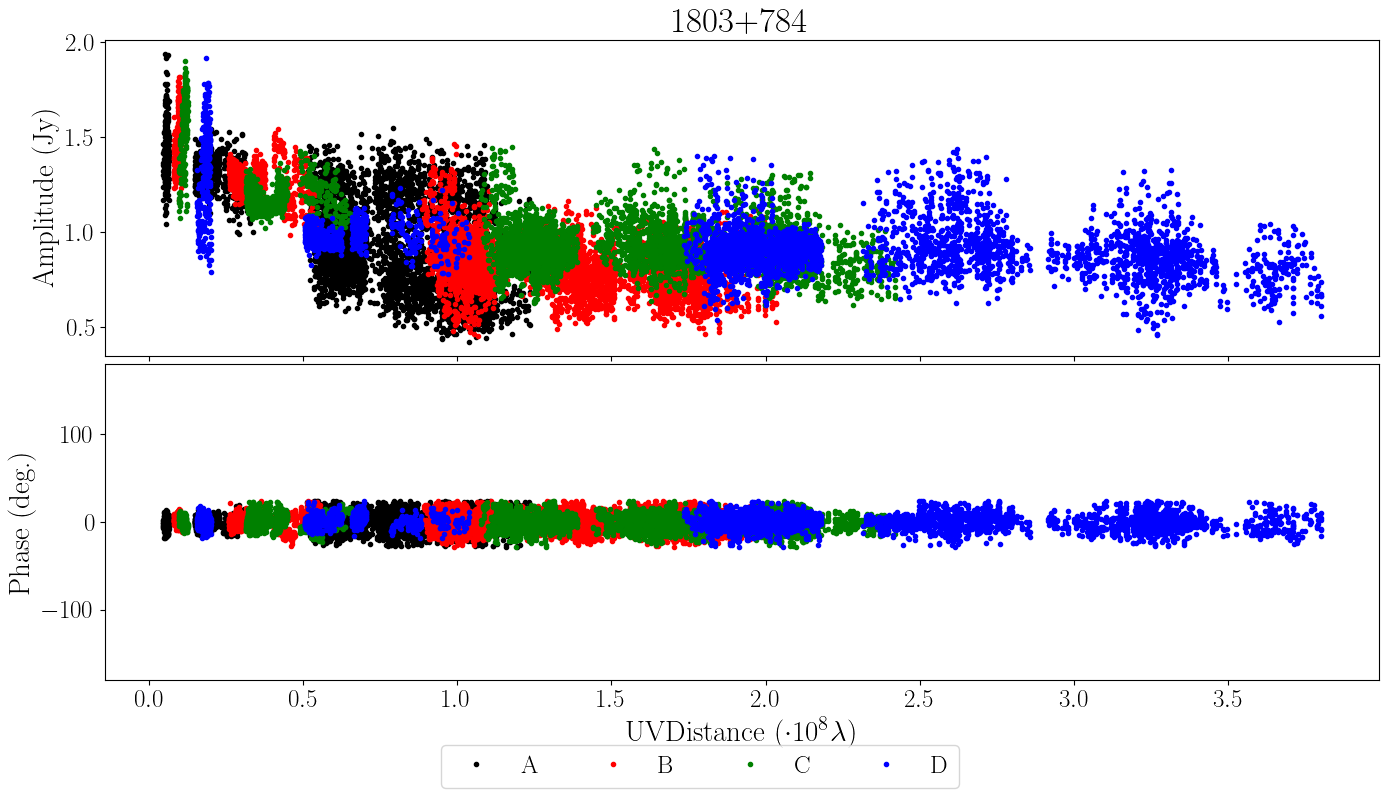}
    \end{minipage}
    \begin{minipage}{0.35\textwidth}
        \centering
        \includegraphics[width=\linewidth]{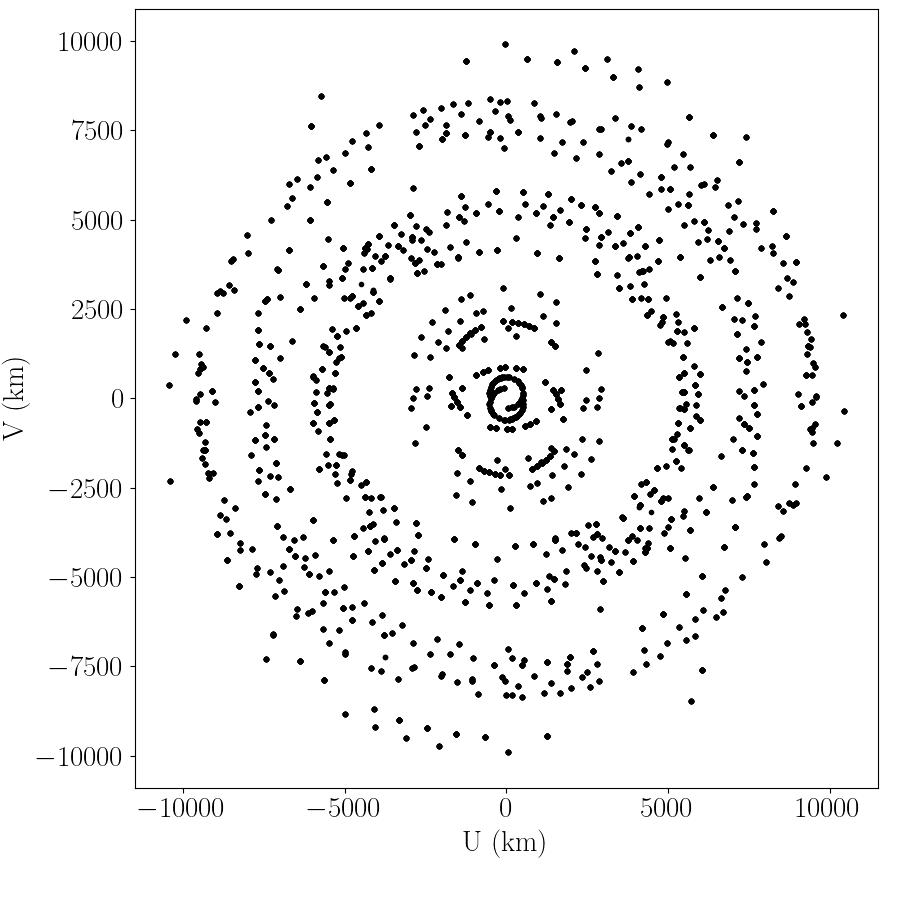}
    \end{minipage}
    \begin{minipage}{0.65\textwidth}
        \centering
        \includegraphics[width=\linewidth]{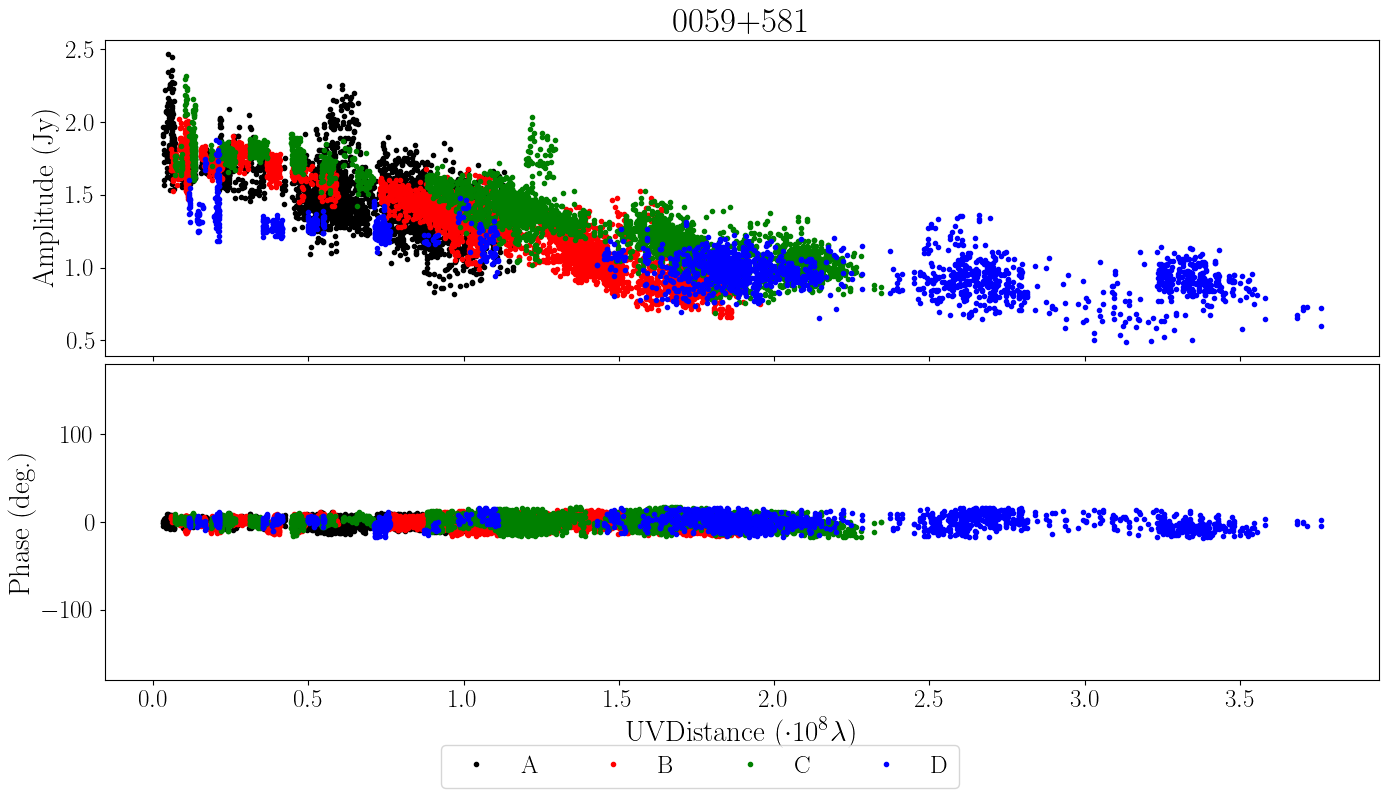}
    \end{minipage}
    \caption{UV-coverage plots for sources 1803+784 (left, top) and 0059+581 (left, bottom). Amplitudes and phases of the RR and LL visibilities as a function of distance in Fourier space (right) in units of wavelength. The different VGOS bands are shown in different colors. 
    }
    \label{fig:UVrad}
\end{figure*}

The data calibration procedure described in this work yields the observed visibilities properly calibrated in phase and amplitude. In Fig. \ref{fig:UVrad}, we show the final results after the phase calibration with the dispersive fringe fitting and the amplitude calibration for two selected sources. We plot the phases and amplitudes of the RR and LL visibilities as a function of the distance in the Fourier plane, showing the expanded UV coverage based on all VGOS bands in different colors. For all bands, we obtain phases close to zero and amplitudes reflecting the structure of our sources, as we show after imaging in the following section. Qualitatively similar results were obtained for the other sources.

\subsection{Imaging of selected sources}
\label{sec:resultsImages}

In the analysis presented in this work, we chose to use regularized maximum likelihood (RML) algorithms for the imaging of the sources, using phase closures, log closure amplitudes, and amplitudes as data. This process simplifies the calibration process because most station-based calibration errors are canceled by using these observables. Furthermore, this method allows us to obtain super-resolution, which favors the study of the effects related to source structure.

Specifically, we used the software ehtim \citep{ehtim2016, ehtim2018}, to which a multifrequency mode for image deconvolution was recently added \citep{ehtimmf}. This method generates a two-term log-log Taylor expansion around a reference image $I_0$ at a reference frequency $\nu_0$,

\begin{equation}
	\log I_i = \log \left(I_0\right)  + \alpha \log \left(\frac{\nu_i}{\nu_0}\right) + \beta \log^2 \left(\frac{\nu_i}{\nu_0}\right) + ...\,,
\end{equation}

where $\alpha$ represents the spectral index map, and $\beta$ is the spectral curvature map.

This method is especially favorable for VGOS data because we have data from four separate frequency bands, allowing us to leverage information from all four bands simultaneously to construct a single intensity-versus-frequency map. Another advantage of this method is the possibility of aligning the images in the different bands to study the structure of the sources, their core shift, and so on. When using closure quantities, absolute spatial information is lost, and imaging the different bands separately therefore does not allow this alignment. This imaging was carried out with an iterative self-calibration process. A more detailed analysis of the complete imaging process will be shown in a following paper.

\begin{figure*}
    \centering
    \subfloat{\includegraphics[trim={0cm 0 0.4cm 0},clip,width=0.33\textwidth]{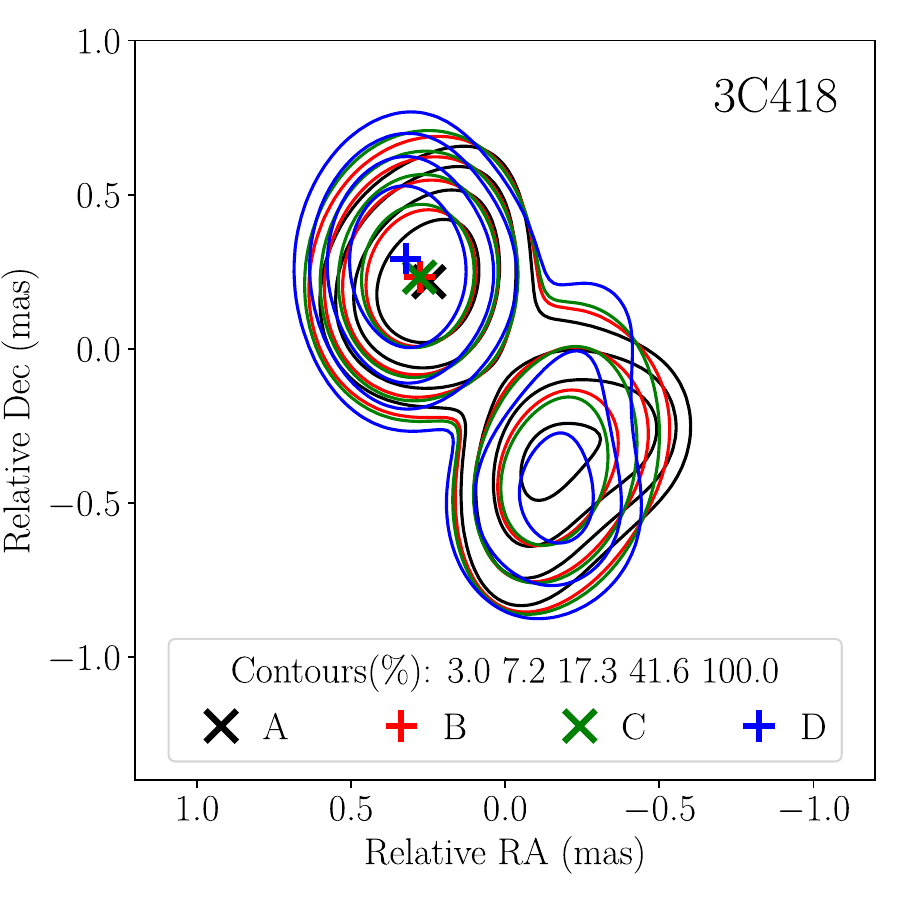}}%
    \subfloat{\includegraphics[trim={0.4cm 0 0 0},clip,width=0.33\textwidth]{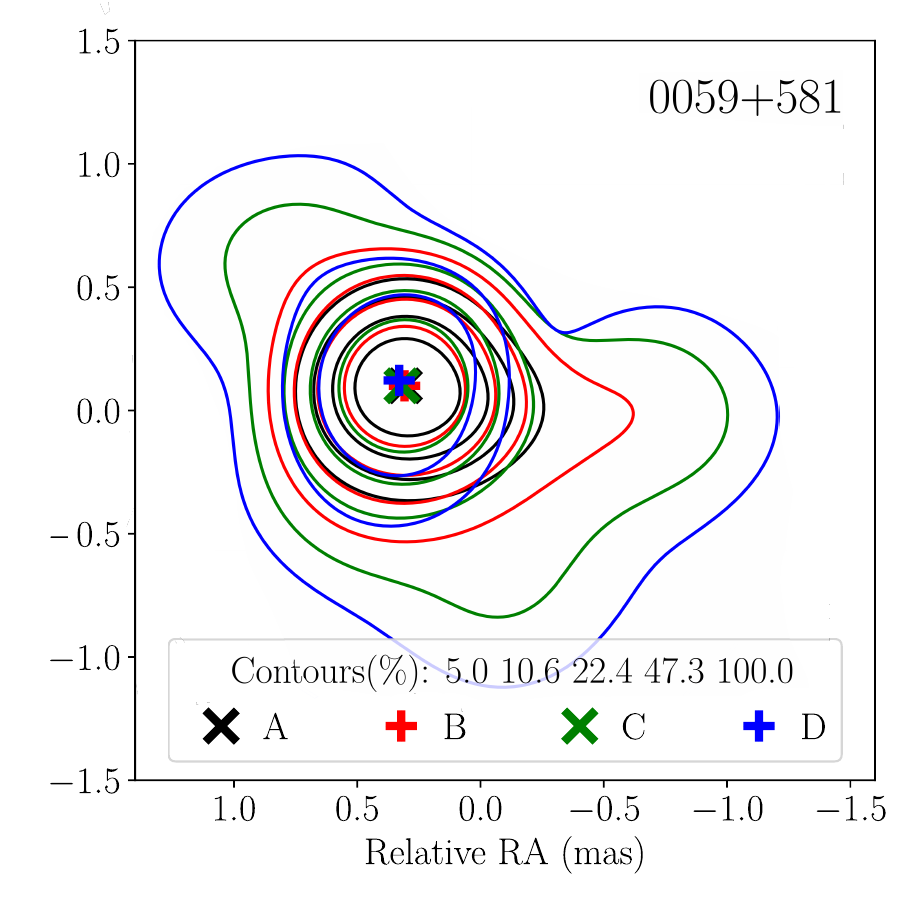}}%
    \subfloat{\includegraphics[trim={0.4cm 0 0 0},clip,width=0.33\textwidth]{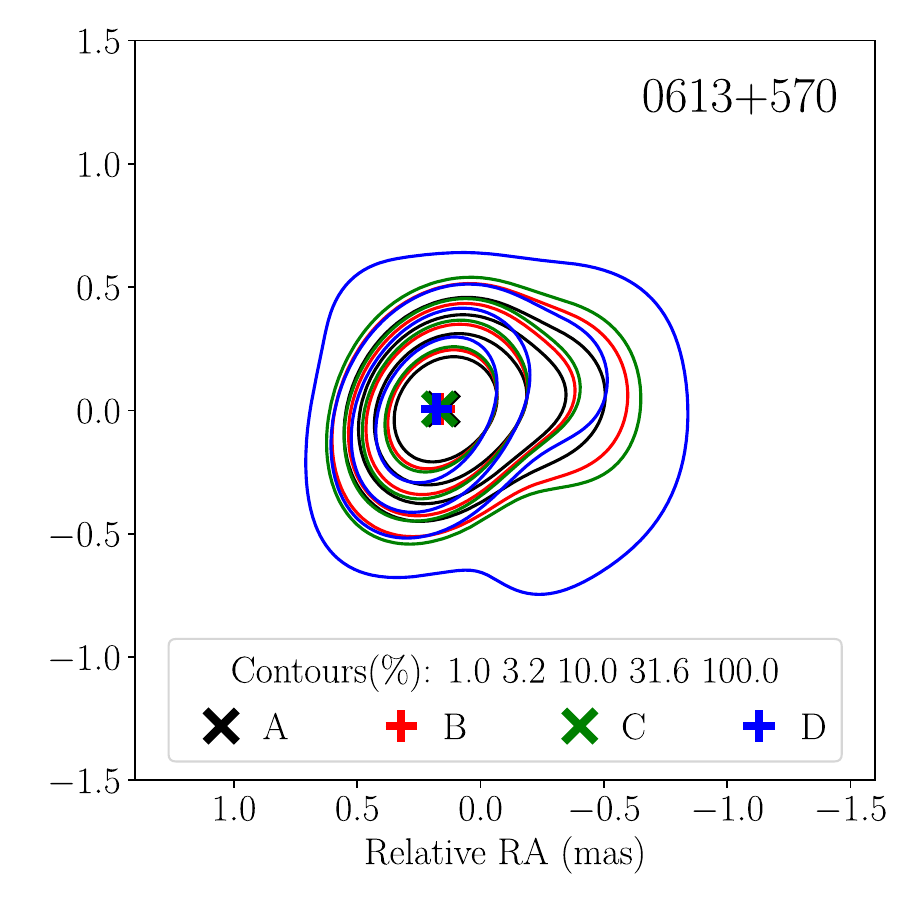}}
    
    \subfloat{\includegraphics[trim={0cm 0 0.4cm 0},clip,width=0.33\textwidth]{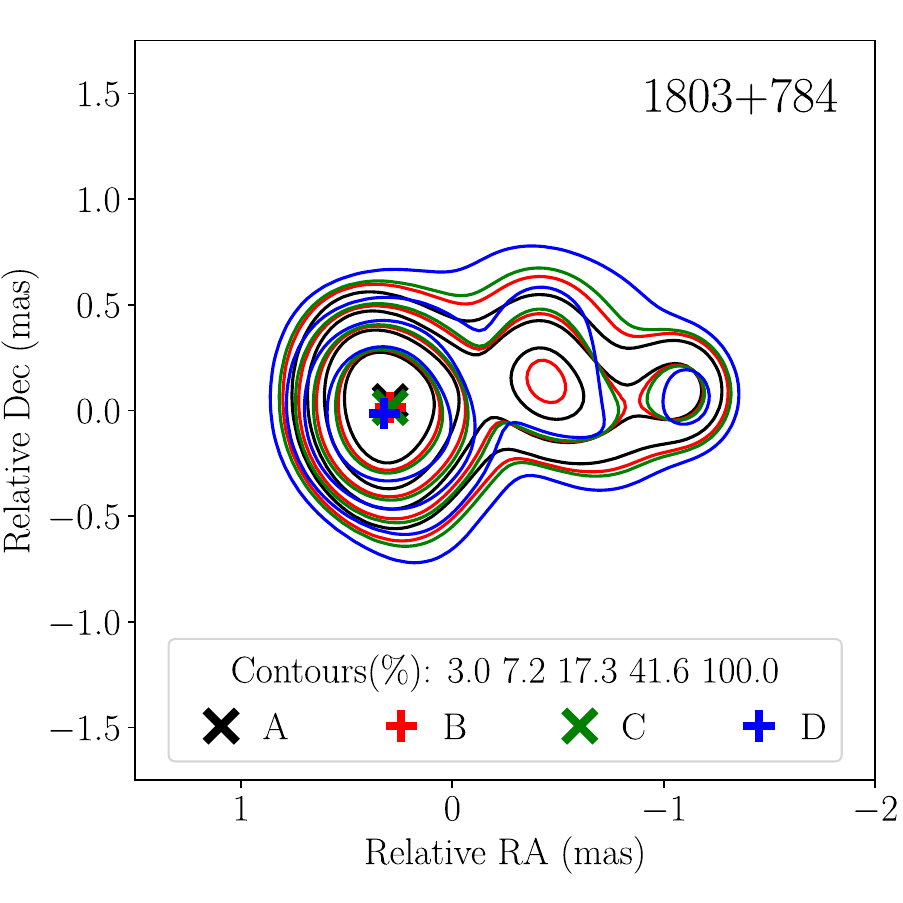}}%
    \subfloat{\includegraphics[trim={0.4cm 0 0 0},clip,width=0.33\textwidth]{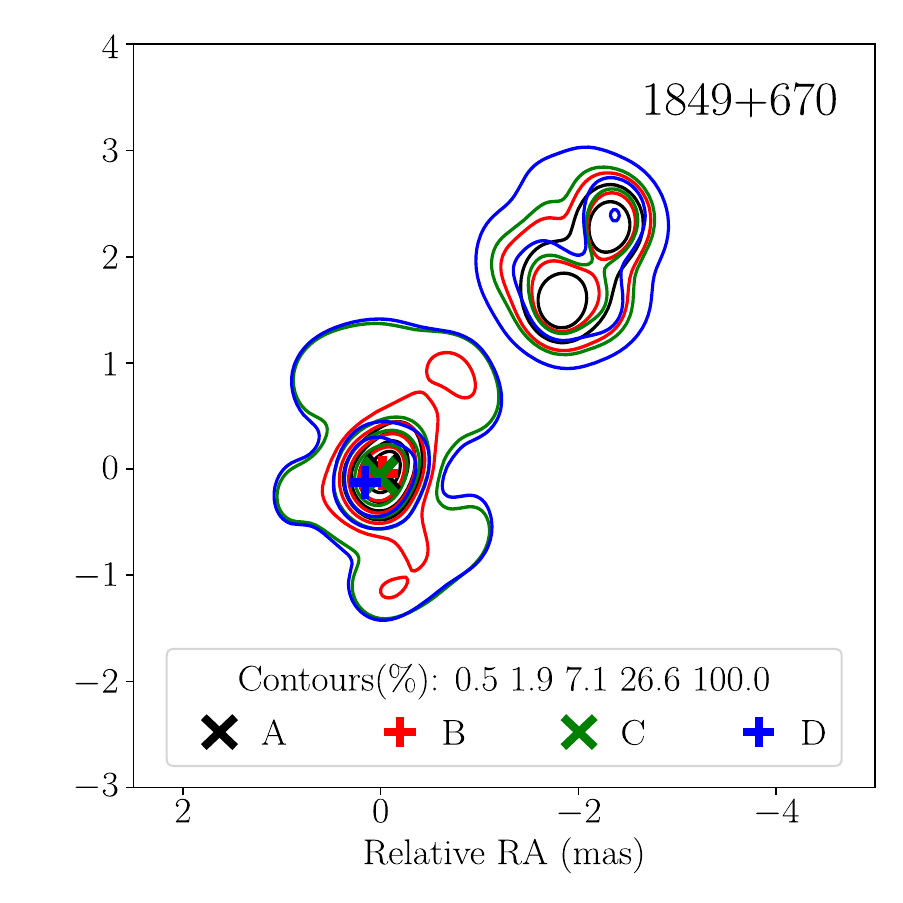}}%
    \subfloat{\includegraphics[trim={0.4cm 0 0 0},clip,width=0.33\textwidth]{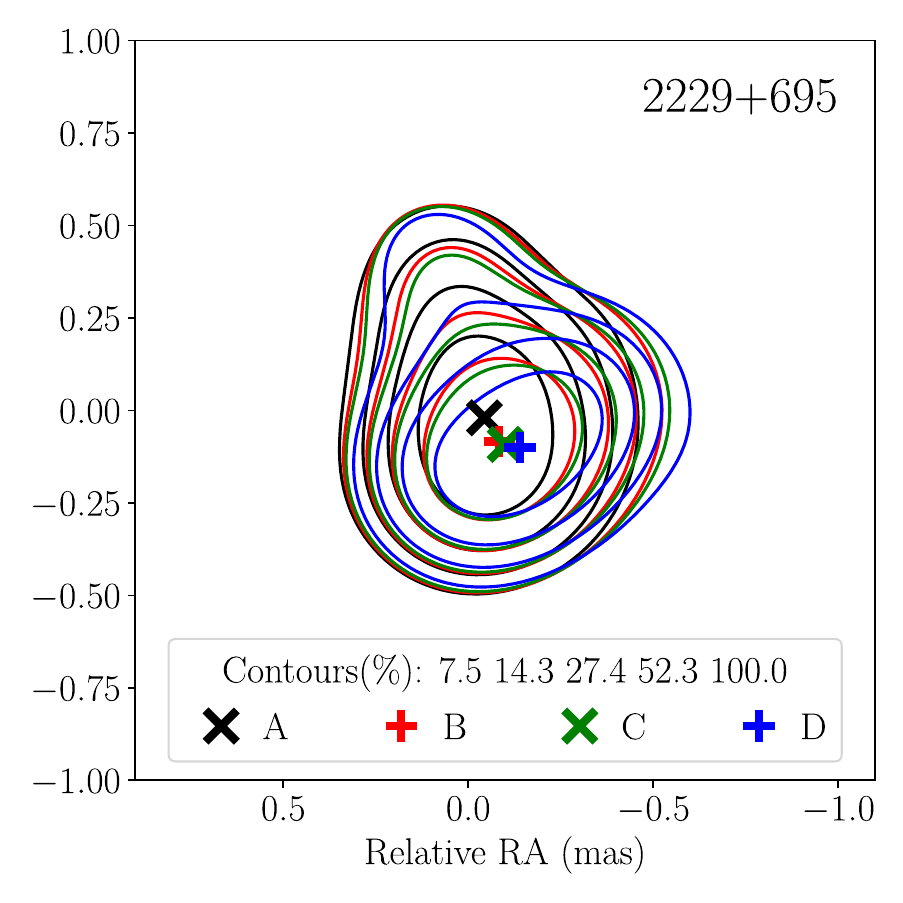}}
    
    \caption{Total intensity maps from VGOS experiment VO2187 obtained from the multifrequency image and the spectral index map. The contours are shown at five levels of the peak percentage, specified in the legend of the plots. Each contour color represents the map for the central frequency of each band: 3.25, 5.5, 6.75, and 10.5 GHz.}
    \label{fig:IMaps}
\end{figure*}

Fig. \ref{fig:IMaps} shows the total intensity maps for the central frequency of each VGOS band, convolved to the same beam to be able to compare them. By displaying the maps at different frequencies together, a study of the core shift of the sources can be made. These images, as well as the direction and distance of the jet, are in line with surveys carried out with other arrays at similar frequencies, such as the images from the MOJAVE program\footnote{https://www.cv.nrao.edu/MOJAVE/}.

\subsection{Full-polarization imaging}

\begin{figure*}
    \centering
  \includegraphics[width=0.99\textwidth]{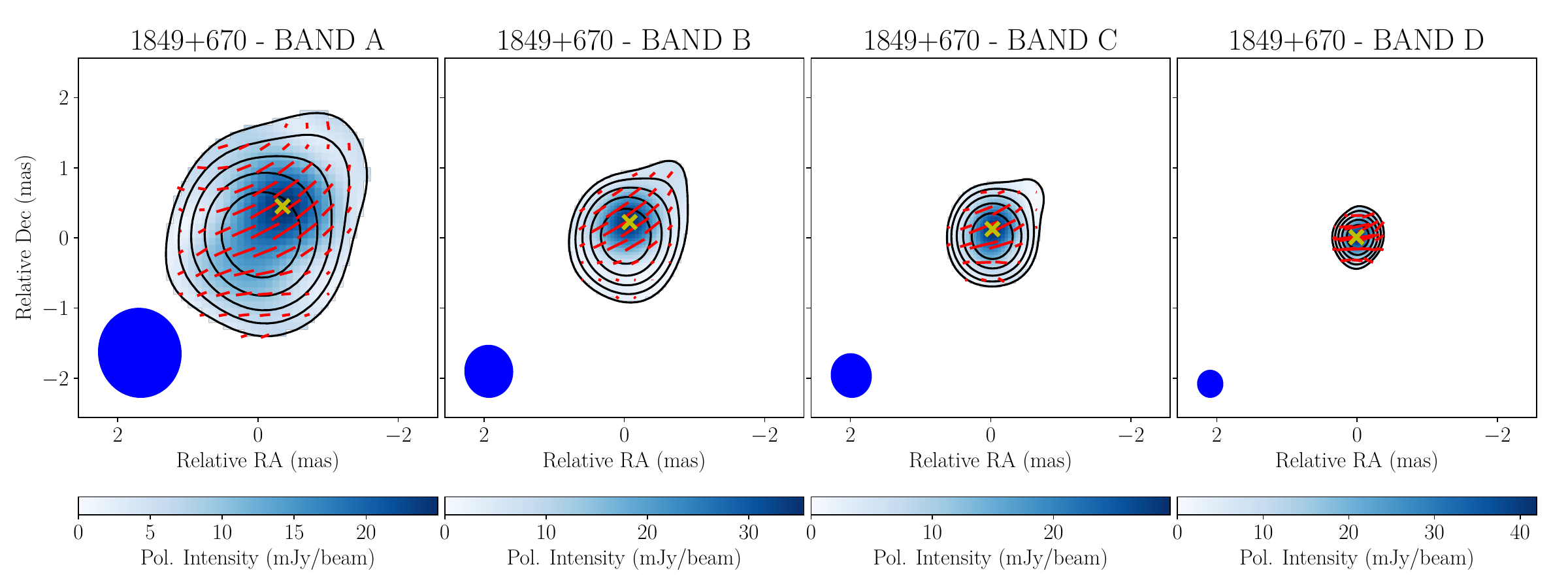}
    \caption{CLEAN full-polarization images of 1849+670 for epoch VO2187, using all the VGOS bands. The FWHM of the convolving beams are shown in the bottom left corners. Contours are shown at five levels of peak percentage in logarithmic scale. The polarization intensity is shown in blue, EVPAs are shown in red, and the polarization peaks are marked as yellow crosses.}
    \label{fig:PMaps}
\end{figure*}

In Fig. \ref{fig:PMaps}, we show full-polarization images of source 1849+670, obtained with the CASA version of CLEAN, for the four VGOS bands. The images were obtained after a D-term calibration with the software \texttt{PolSolve} \citep{PolSolve}. These Dterms correct for residual cross-polarization gains after the polconversion \citep[e.g., ][]{GoddiAPP} and will be discussed in a forthcoming publication.

In Fig. \ref{fig:PMaps}, the total-intensity distributions (black contours) show a jet extension in the northwest direction, the same as the jet in the images published as part of the MOJAVE program \citep{MOJAVE}. The position of the polarization intensity peaks (shown as yellow crosses) is also aligned in the jet direction with respect to the total-intensity peaks, although a hint of counterclockwise rotation can be seen from bands A to C (the polarization peak in band D coincides with the total-intensity peak at the CLEAN resolution). Curiously enough, the expected core shift should move the total-intensity core southeast at higher frequencies (i.e., the direction opposite to the jet stream). However, the polarization peak approaches the total-intensity peak at higher frequencies. If the polarization peak were related to an optically thin jet feature, we should observe the opposite relative astrometry between total intensity and polarization: The distance between polarization and total-intensity peaks should increase with increasing frequency. The observed behavior of the polarization peak can only be explained if the region that emits the polarized emission is also optically thick.

Astrometry in the different VGOS bands is necessary to obtain more precise geodetic observables because the core shift breaks the proportionality relation between the interferometric phase and the group delay \citep{Porcas2009}. The only way to accurately obtain multifrequency astrometry of the sources (i.e., the absolute coordinates of the source at each frequency) would be to incorporate phase-referencing observations to monitor the core shift at VGOS frequencies. We lack observations like this in this experiment. The best way to align (in relative coordinates) the images in different bands therefore is to create multifrequency models aligning the jet because they must be optically very thin.

The EVPAs (from north to west) at the polarization intensity peaks are 56.3, 56.7, 67.5, and 81.2 degrees for bands A to D, respectively. These results do not follow the $\lambda^2$ relation expected for Faraday-thin external rotation-measure screens, which is another indication of optical thickness in the region producing the polarized emission. A more detailed discussion of the polarization properties of this and other sources observed in the VO2187 experiment will be published elsewhere.


\subsection{Proposed strategies to improve the VGOS data quality}
\label{sec:resultsStrategies}

VGOS observations are scheduled to maximize figures of merit related to the geodetic precision. The exploitation of VGOS visibilities for astronomical use is thus suboptimal. However, with minimum changes in the observation schedule, it is possible to improve the astronomical capabilities of VGOS and to use them in turn to optimize the geodetic precision by accounting for the frequency-dependent source structure and polarization. Based on the results reported here, we suggest several actions that would help to improve the quality of the VGOS calibration for both geodetic and astronomical use.

\begin{enumerate}

\item The inclusion of long (1-2 minute) calibration scans a few times during an observation would allow us to determine (and track) the cross-polarization gains with a high precision. This would result in a polconversion with a minimum instrumental polarization.

\item The frequency separation of the phase-cal tones should be as small as possible to ensure a robust and precise estimate of the instrumental delays within each spw. This is especially needed for the Yebes station, where the 10\,MHz separation may prevent a correct delay estimate if a small fraction of the tone phases cannot be determined.

\item Phase-cal tones at twin (i.e., close by) stations should be offset one from the other in order to avoid spurious cross-correlations that degrade the quality of the resulting intra-site fringes. Adding the baselines between twin stations would add more robustness to the global (antenna-based) calibration of gains and instrumental polarization.

\item Gain curves and system temperatures should be characterized and monitored, respectively, to allow for the amplitude calibration of the visibilities. With this information and the GFF solutions, it would easily be possible to construct full-polarization wide-band images of the observed sources.

\item Phase-referencing observations should be added throughout the VGOS experiments in order to subtract the influence of core shift on group-delay measurements.

\end{enumerate}





\section{Conclusions}
\label{sec:conclusions}

The linear polarization basis is preferably used in new-generation radio interferometers from ALMA and the EHT to some stations of the EVN and the whole VGOS. The main reason is the wide instantaneous fractional bandwidths that can be achieved with a minimum instrumental polarization. However, for the case of VLBI, the use of linear polarizers hamper the full-polarization calibration through the different parallactic angles of the interferometer elements.

VGOS is a global VLBI array that only observes in the basis of linear polarization. We have successfully used the algorithm PolConvert \citep{PolConvert} to recompute the visibilities of a VGOS epoch (VO\,2187, observed in July 2022) in the basis of circular polarization by estimating the cross-polarization gains of all the antennas and using this information for the conversion. Then, we fully processed and calibrated the polconverted data. To our knowledge, this is the first time that full-polarization information was extracted for sources observed with a VLBI array that only uses linear polarizers. 

We compared the estimated cross-polarization gains of a subset of antennas to those derived from another VGOS experiment obtained two months later \citep{Jaron23}. The differences of cross-delays between polarizers only deviate by a few picoseconds between these two epochs, with the exception of the Yebes station (where a difference of about 10\,ps is found), likely related to updates in the instrumental phase-calibration system at this station.

The polarization conversion was assessed by comparing the fringe amplitudes of the parallel-hand products (RR and LL) to those of the cross-hand correlations (RL and LR), as well as the parallel-hand phases as a function of parallactic-angle difference among antennas. All tests indicate a successful conversion.

We have also developed a new GWBFF algorithm\footnote{The pipeline is available at github.com/marti-vidal-i/PolConvert}, which we presented in this report. The algorithm subtracts an a priori model of the ionospheric dispersive delays by using IONEX maps and applies a novel approach for a fast estimate of dispersive and nondispersive antenna-based delays and phases. We applied our algorithm to the polconverted VGOS observations and compared the calibrated visibility phases to those obtained with another wideband global fringe fitter (the one implemented in the CASA software package by NRAO; see Appendix \ref{CASAComparison}). From this comparison, we find that (at least for the data reported here) our algorithm rutinely produces shorter calibrated phases in absolute value.  

Then, we applied an amplitude calibration based on an incomplete set of gain curves and system temperatures for a subset of the VGOS stations. This is the first time that this type of calibration has been applied to VGOS observations and represents the first step toward a complete satisfactory amplitude calibration in geodetic experiments.

Finally, we generated full-polarization wide-band model images for a subset of the most frequently observed sources in the VO\,2187 epoch. With the GWBFF, this is the first time that full-polarization and multifrequency images were obtained from sources observed with VGOS. The results are compatible with VLBI images of the same sources produced with different (astronomical) VLBI arrays.

This work is the first in a series, where the use of GFF on polconverted visibilities are compared to the products obtained from the official (pseudo-Stokes I) calibration approach used by the IVS, and information about the jet structures of the observed AGN and the magnet-ionic medium around them is retrieved.


 \begin{acknowledgements}
This work has been partially supported by the Generalitat Valenciana GenT Project CIDEGENT/2018/021 and by the MICINN Research Project PID2019-108995GB-C22.

This work has been supported by the grant PRE2020-092200 funded by MCIN/AEI/ 10.13039/501100011033 and by ESF invest in your future.

We acknowledge support from the Astrophysics and High Energy Physics programme by MCIN, with funding from European Union NextGenerationEU (PRTR-C17I1) and the Generalitat Valenciana through grant ASFAE/2022/018.

We thank Bill Petrachenko for his work in estimating the flux density as a function of uv distance in order to perform the amplitude calibration.

We also thank Eduardo Ros from the MPIfR for reading the manuscript and for giving us valuable advice to improve it.

\end{acknowledgements}

%
%

\bibliographystyle{aa}
\bibliography{vo2187.bib}

\begin{appendix}
\section{Absolute EVPA calibration}
\label{sec:EVPA}

In an ordinary VLBI observation using circular-polarization receivers, any instrumental phase offset between the $R$ and $L$ polarization channels, uniformly added to all the VLBI antennas of the interferometer, cannot be distinguished from a global rotation of the electric vector position angle (EVPA) of a linearly polarized source. This limitation in the EVPA calibration using circularly-polarized antenna feeds has traditionally been a major issue in VLBI polarimetric studies. 

Usually, in full-polarization VLBI, the use of data from other arrays (like the VLA) was required at similar epochs, in order to force a match between the EVPAs observed with VLBI and those from the other array. Needless to say that such an EVPA calibration approach suffers from important potential biases, related to the extended (polarized) scales that may affect the EVPAs measured with, e.g., the VLA, that may be completely resolved out at the VLBI resolution. 

However, as discussed in \cite{PolConvert} and \cite{GoddiAPP}, an important advantage of the use of linear polarizers in VLBI, via the algorithm PolConvert, is that the absolute phase between the polconverted $R$ and $L$ polarizations (i.e., in essence, the absolute EVPA) is automatically calibrated, as long as a polconverted antenna is used as the reference station in the process of GFF. In the case of VGOS, all antennas have linear-polarization feeds (i.e., all of them are being polconverted), so the condition for an absolute EVPA calibration is always satisfied.

Having the phases between $R$ and $L$ absolutely calibrated allows us to combine the $RR$ and $LL$ visibilities right after the polconversion, so that the fringes corresponding to Stokes I can be obtained. This improves the S/R by $\sqrt{2}$, as compared to the independent processing of $RR$ and $LL$.
However, for a correct combination of $RR$ and $LL$, the different orientations of the two antenna feeds (i.e., basically, the difference between the parallactic angles of the antennas) have to be accounted as well. Once the parallactic angle effects are removed from $RR$ and $LL$, both visibility products should have the same phases, assuming that Stokes V is negligible compared to Stokes I. 

We have tested this statement by studying the phases of the $RR/LL$ visibility ratio as a function of frequency (i.e., spw) and time. In order to be sensitive to the (small) potential deviations in the phases of the $RR/LL$ ratio, we have restricted the analysis to the scans with the highest S/R; in particular, we have selected those scans with S/R$>$10 in all spw (only 120 scans out of 1919 passed this restriction). 

\begin{figure}[ht!]
    \centering
    \includegraphics[width=0.5\textwidth]{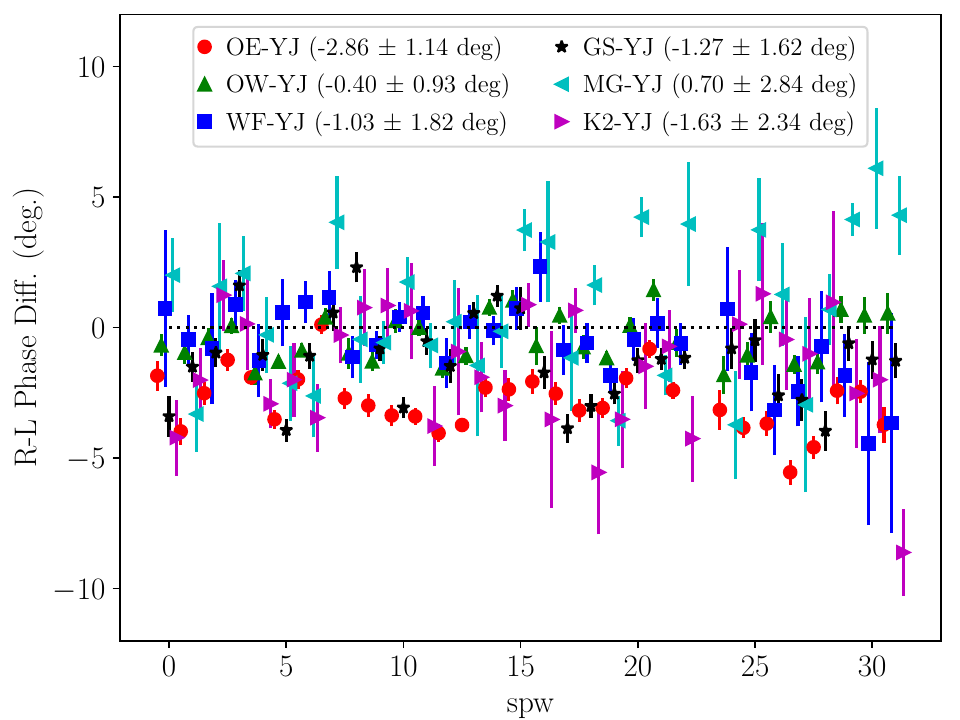}
    \caption{Scan average of the $RR/LL$ phase for all spw, using the scans with highest S/R (see text) for the baselines to YJ. The spw 23 has been removed from the analysis. Different baselines are slightly shifted in the spw axis, for clarity. The complete statistics for each baseline, averaged over spw, are included in the figure labels.}
    \label{fig:RLDiff}
\end{figure}

In Fig. \ref{fig:RLDiff}, we show the phases of $RR/LL$ (after accounting for the parallactic angles), averaged over all the detections with S/R$>$10 and referenced to the YJ antenna. The error bars shown in the figure correspond to the scatter of the $RR/LL$ phases across the scans (which involve different sources under different parallactic angles). It is remarkable the very little scatter in the $RR/LL$ phases across scans of baseline OE-YJ (error bars of the order of only one degree), which happens to be a relatively short baseline. Actually, for that baseline, the average $RR/LL$ phase over scans and spw is different from zero at a $\sim2-3\sigma$ level ($-2.86\pm1.14$\,deg). The unexpected (though small) phase offset between $RR$ and $LL$ in this baseline could be explained, for instance, as due to a small rotation of the receiver feeds of these antennas with respect to the horizontal axis of their mounts.

For the rest of the baselines, all $RR/LL$ phases are compatible with zero and take typical values lower than 10 degrees, which is indicative of a proper conversion of the data into a circular polarization basis. Otherwise, the phases of $RR$ and $LL$ would not be rotating according to the parallactic angle of each scan, and the parallactic-angle-corrected phases of the $RR/LL$ ratios would not be compatible with zero.


\section{Comparison with CASA fringe fitting}

\begin{figure*}[b!]
  \centering
  \includegraphics[width=0.995\textwidth]{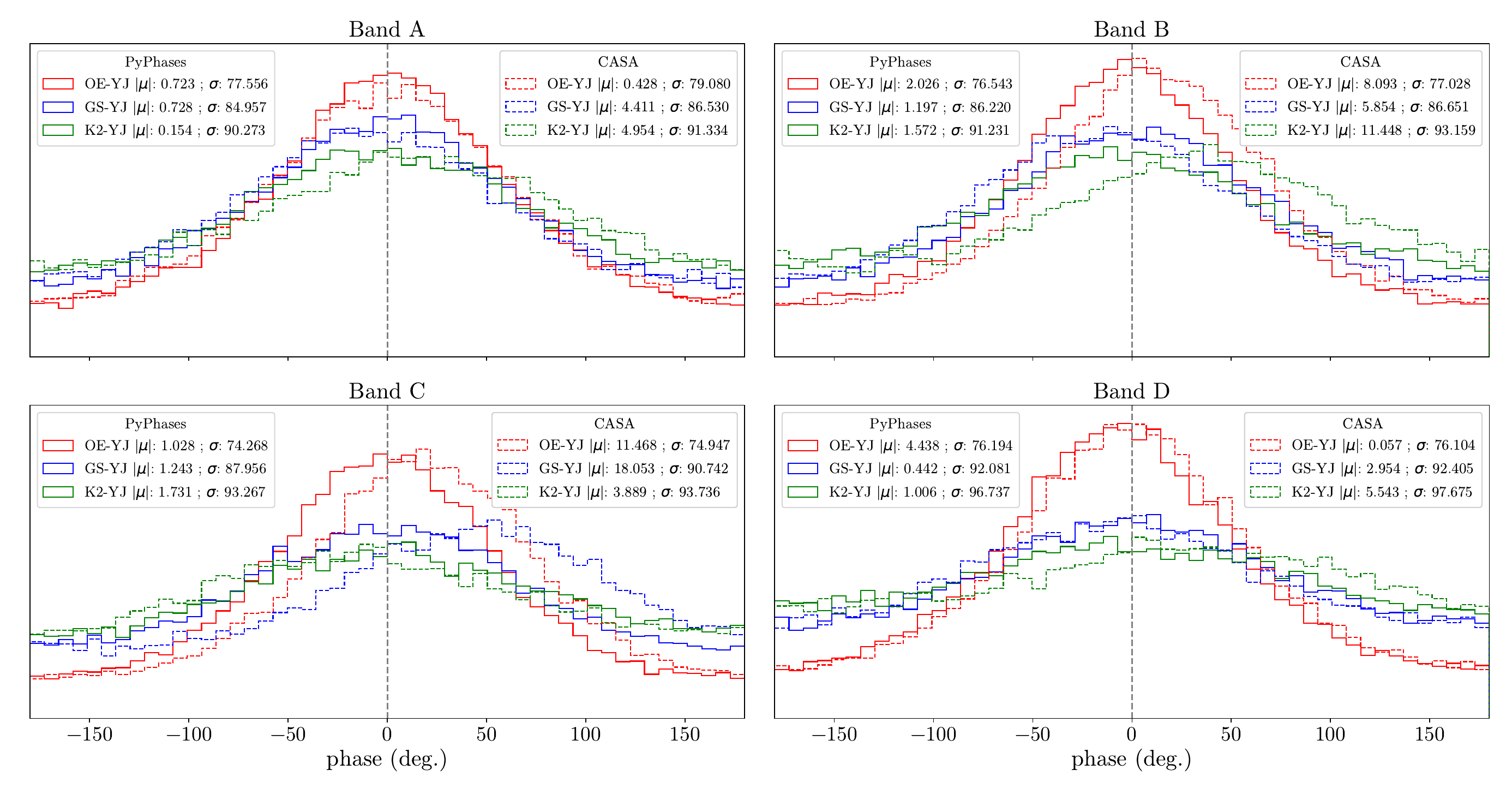}
  \caption{Phase histograms of the visibilities of scan 1362 (observed at 10:50\,UT), of the source 0059+581, for a selection of baselines, after the Global Fringe Fitting calibration. Solid lines, using our WBGFF algorithm implemented in PyPhases; dotted lines, using the implementation in CASA (see text). In the boxes, we show the absolute value of the mean (\(|\mu|\)) in deg. and the standard deviation (\(\sigma\)) in deg. for each distribution.
    \label{fig:GFFvsCASA}} 
\end{figure*}

The performance of our WBGFF is also comparable to (and, for the data reported here, slightly better than) other WBGFF implementations, like the one included in the latest versions of the CASA software \citep{CASARef}. As an example, Fig. \ref{fig:GFFvsCASA} shows phase histograms (after the WBGFF calibration) of the scan observed at 10:50\,UT (source 0059$+$581, which is almost point-like; see Sect. \ref{sec:resultsImages}), for a selection of baselines. In principle, all phase distributions should peak at values as close to zero as possible (as long as the least-squares minimization has converged successfully) and have standard deviations related to the S/R of each baseline.

For the case of our WBGFF, implemented in PyPhases, we find that the phase distributions are usually more centered around zero than the corresponding distributions obtained with the CASA implementation. This can be seen quantitatively by looking at the mean and standard deviation of the phase distributions (shown in the legends of Fig. \ref{fig:GFFvsCASA}). Some extreme cases are seen, for instance, in Band C, where the GS-YJ has an average phase of 18\,deg. (and a peak around 50\,deg.) for the CASA case, whereas it takes a value around only 1\,deg. for the PyPhases case.

The usual method in CASA also consists of carrying out a manual phase-cal, in which they calculate the single-band delay for each spw. However, their fringe fitting works differently, starting with a 2D Fast Fourier Transform that does not include dispersion and subsequently performing a least squares minimization stage that optionally includes a dispersion term. Furthermore, no a priori information is included in CASA. After applying the corrections from the manual phase-cal, the delay is calculated, either single-band delay (each spw separately) or combining every spw for calculating a multiband delay, without combining both types of delays \citep{vanBemmel2022}.
\label{CASAComparison}

\end{appendix}

\end{document}